\documentstyle[bezier]{article}
\catcode`\@=11
\ifcase\@ptsize
 \font\tenmsa=msam10
 \font\sevenmsa=msam7
 \font\fivemsa=msam5
 \font\tenmsb=msbm10
 \font\sevenmsb=msbm7
 \font\fivemsb=msbm5
 \font\teneu=eufm10
 \font\seveneu=eufm7
 \font\fiveeu=eufm5
 \font\tenib=cmmib10
 \font\sevenib=cmmib7
 \font\fiveib=cmmib5
\or
 \font\tenmsa=msam10 scaled \magstephalf
 \font\sevenmsa=msam7 scaled \magstephalf
 \font\fivemsa=msam5 scaled \magstephalf
 \font\tenmsb=msbm10 scaled \magstephalf
 \font\sevenmsb=msbm7 scaled \magstephalf
 \font\fivemsb=msbm5  scaled \magstephalf
 \font\teneu=eufm10  scaled \magstephalf
 \font\seveneu=eufm7  scaled \magstephalf
 \font\fiveeu=eufm5   scaled \magstephalf
 \font\tenib=cmmib10  scaled \magstephalf
 \font\sevenib=cmmib7  scaled \magstephalf
 \font\fiveib=cmmib5   scaled \magstephalf
\or
 \font\tenmsa=msam10 scaled \magstep1
 \font\sevenmsa=msam7 scaled \magstep1
 \font\fivemsa=msam5  scaled \magstep1
 \font\tenmsb=msbm10 scaled \magstep1
 \font\sevenmsb=msbm7 scaled \magstep1
 \font\fivemsb=msbm5  scaled \magstep1
 \font\teneu=eufm10   scaled \magstep1
 \font\seveneu=eufm7 scaled \magstep1
 \font\fiveeu=eufm5 scaled \magstep1
 \font\tenib=cmmib10     scaled \magstep1
 \font\sevenib=cmmib7   scaled \magstep1
 \font\fiveib=cmmib5   scaled \magstep1
\fi
\newfam\msafam
\textfont\msafam=\tenmsa  \scriptfont\msafam=\sevenmsa
  \scriptscriptfont\msafam=\fivemsa
\newfam\msbfam
\textfont\msbfam=\tenmsb  \scriptfont\msbfam=\sevenmsb
  \scriptscriptfont\msbfam=\fivemsb
\def\Bbb{\ifmmode\let\next\Bbb@\else
 \def\next{\errmessage{Use \string\Bbb\space only in math mode}}\fi\next}
\def\Bbb@#1{{\Bbb@@{#1}}}
\def\Bbb@@#1{\fam\msbfam#1}
\newfam\eufam
\textfont\eufam=\teneu  \scriptfont\eufam=\seveneu
  \scriptscriptfont\eufam=\fiveeu
\def\frak{\ifmmode\let\next\frak@\else
 \def\next{\errmessage{Use \string\frak\space only in math mode}}\fi\next}
\def\frak@#1{{\frak@@{#1}}}
\def\frak@@#1{\fam\eufam#1}
\newfam\ibfam
\textfont\ibfam=\tenib  \scriptfont\ibfam=\sevenib
  \scriptscriptfont\ibfam=\fiveib
\def\bold{\ifmmode\let\next\bold@\else
 \def\next{\errmessage{Use \string\bold\space only in math mode}}\fi\next}
\def\bold@#1{{\bold@@{#1}}}
\def\bold@@#1{\fam\ibfam#1}
\def\hexnumber@#1{\ifcase#1 0\or 1\or 2\or 3\or 4\or 5\or 6\or 7\or 8\or
 9\or A\or B\or C\or D\or E\or F\fi}
\def\newsymbolb#1#2#3#4{\mathchardef#1="#2\hexnumber@\msbfam#3#4}
\def\newsymbola#1#2#3#4{\mathchardef#1="#2\hexnumber@\msafam#3#4}
\newsymbola{\leqslant}336
\newsymbola{\geqslant}33E
\newsymbola{\square}003
\newsymbola{\lessgtr}337
\newsymbola{\gtrless}33F
\newsymbola{\ravnodots}33A
 \let\leq\leqslant
 \let\geq\geqslant

\def\hybrid{\topmargin 0pt      \oddsidemargin 0pt
        \headheight 0pt \headsep 0pt
        \textwidth 165true mm       
        \textheight 231true mm         
        \marginparwidth 0.0in
        \parskip 0pt plus 1pt   \jot = 1.5ex}
\catcode`\@=11
\def\marginnote#1{}

\newcount\hour
\newcount\minute
\newtoks\amorpm
\hour=\time\divide\hour by60
\minute=\time{\multiply\hour by60 \global\advance\minute by-\hour}
\edef\standardtime{{\ifnum\hour<12 \global\amorpm={am}%
        \else\global\amorpm={pm}\advance\hour by-12 \fi
        \ifnum\hour=0 \hour=12 \fi
        \number\hour:\ifnum\minute<10 0\fi\number\minute\the\amorpm}}
\edef\militarytime{\number\hour:\ifnum\minute<10 0\fi\number\minute}

\def\draftlabel#1{{\@bsphack\if@filesw {\let\thepage\relax
   \xdef\@gtempa{\write\@auxout{\string
      \newlabel{#1}{{\@currentlabel}{\thepage}}}}}\@gtempa
   \if@nobreak \ifvmode\nobreak\fi\fi\fi\@esphack}
        \gdef\@eqnlabel{#1}}
\def\@eqnlabel{}
\def\@vacuum{}
\def\draftmarginnote#1{\marginpar{\raggedright\scriptsize\tt #1}}

\def\draft{\oddsidemargin -.5truein
        \def\@oddfoot{\sl preliminary draft\quad\filename \hfil
        \rm\thepage\hfil\sl\today\quad\militarytime}
        \let\@evenfoot\@oddfoot \overfullrule 3pt
        \let\label=\draftlabel
        \let\marginnote=\draftmarginnote
   \def\@eqnnum{(\theequation)\rlap{\kern\marginparsep\tt\@eqnlabel}%
\global\let\@eqnlabel\@vacuum}  }
\newcounter{app}
\newcounter{sapp}[app]
\def\theapp{\Alph{app}}
\newcommand{\app}[1]{
\refstepcounter{app}{\vspace{7mm}
\noindent\Large\bf Appendix
\theapp.
 \ #1 \par \vspace{5mm}}
\setcounter{equation}{0}
\def\theequation{\Alph{app}.\arabic{equation}}}

\makeatletter
\newdimen\normalarrayskip              
\newdimen\minarrayskip                 
\normalarrayskip\baselineskip
\minarrayskip\jot
\newif\ifold             \oldtrue            
\def\arraymode{\ifold\relax\else\displaystyle\fi} 
\def\eqnumphantom{\phantom{(\theequation)}}     
\def\@arrayskip{\ifold\baselineskip\z@\lineskip\z@
     \else
     \baselineskip\minarrayskip\lineskip2\minarrayskip\fi}
\def\@arrayclassz{\ifcase \@lastchclass \@acolampacol \or
\@ampacol \or \or \or \@addamp \or
   \@acolampacol \or \@firstampfalse \@acol \fi
\edef\@preamble{\@preamble
  \ifcase \@chnum
     \hfil$\relax\arraymode\@sharp$\hfil
     \or $\relax\arraymode\@sharp$\hfil
     \or \hfil$\relax\arraymode\@sharp$\fi}}
\def\@array[#1]#2{\setbox\@arstrutbox=\hbox{\vrule
     height\arraystretch \ht\strutbox
     depth\arraystretch \dp\strutbox
     width\z@}\@mkpream{#2}\edef\@preamble{\halign \noexpand\@halignto
\bgroup \tabskip\z@ \@arstrut \@preamble \tabskip\z@ \cr}%
\let\@startpbox\@@startpbox \let\@endpbox\@@endpbox
  \if #1t\vtop \else \if#1b\vbox \else \vcenter \fi\fi
  \bgroup \let\par\relax
  \let\@sharp##\let\protect\relax
  \@arrayskip\@preamble}
\def\eqnarray{\stepcounter{equation}%
              \let\@currentlabel=\theequation
              \global\@eqnswtrue
              \global\@eqcnt\z@
              \tabskip\@centering
              \let\\=\@eqncr
              $$%
 \halign to \displaywidth\bgroup
    \eqnumphantom\@eqnsel\hskip\@centering
    $\displaystyle \tabskip\z@ {##}$%
    &\global\@eqcnt\@ne \hskip 2\arraycolsep
         $\displaystyle\arraymode{##}$\hfil
    &\global\@eqcnt\tw@ \hskip 2\arraycolsep
         $\displaystyle\tabskip\z@{##}$\hfil
         \tabskip\@centering
    &{##}\tabskip\z@\cr}
\makeatother

\relax

\hybrid

\def\bar{\overline}
\def\*{\star}
\def\[{\left[}
\def\]{\right]}
\def\({\left(}      
\def\){\right)}

\def\frac#1#2{{#1 \over #2}}

\def\d{\partial}

\def\lvac{\hbox{$\langle 0 \vert $}}
\def\2pi{\hbox{$2\pi i$}}

\def\dsl{\raise.15ex\hbox{/}\kern-.57em\partial}
\def\Dsl{\,\raise.15ex\hbox{/}\mkern-.13.5mu D}
\def\th{\theta}     
     \def\Ga{\Gamma}

\def\al{\alpha}
\def\ep{\epsilon}
\def\la{\lambda}

      \def\CC{{\cal C}}
      
   \def\CH{{\cal H}}   
      
      \def\CO{{\cal O}}
      \def\CR{{\cal R}}

\def\lvac{\hbox{$\langle 0 \vert $}}

\def\2pi{\hbox{$2\pi i$}}

\begin{document}
\def\filename{ang.tex}
\def\bea{\begin{eqnarray}}
\def\eea{\end{eqnarray}}
\def\beq{\begin{equation}}          \def\bn{\beq}
\def\eeq{\end{equation}}            \def\ed{\eeq}
\def\nn{\nonumber}                  \def\g{\gamma}
\def\ot{\otimes}
\def\sk#1{\left({#1}\right)}
\def\id{{\rm id}}
\def\tr{{\rm tr}}
\def\tah{{\rm th}}
\def\sh{{\rm sh}}
\def\ch{{\rm ch}}
\def\ctg{{\rm ctg}}
\def\cth{{\rm cth}}
\def\tg{{\rm tg}}
\def\th{{\rm th}}
\def\sign{{\rm sign}}
\def\qdet{{\rm q-det}}
\def\Re{{\rm Re}\,}
\def\Im{{\rm Im}\,}
\def\RR{\Bbb{R}}
\def\ZZ{\Bbb{Z}}
\def\CC{\Bbb{C}}
\def\PP{\Bbb{P}}
\def\r#1{\mbox{(}\ref{#1}\mbox{)}}
\def\ep{\varepsilon}
\def\eps{\epsilon}
\def\ve{\ep}
\def\fract#1#2{{\mbox{\footnotesize $#1$}\over\mbox{\footnotesize $#2$}}}
\def\stackreb#1#2{\ \mathrel{\mathop{#1}\limits_{#2}}}
\def\res#1{\stackreb{\mbox{\rm res}}{#1}}
\def\lim#1{\stackreb{\mbox{\rm lim}}{#1}}
\def\Res#1{\stackreb{\mbox{\rm Res}}{#1}}
\let\dis=\displaystyle
\def\ee{{\bf e}}
\def\ff{{\bf f}}
\def\hh{{\bf h}}
\def\D{\Delta}
\def\HH{{\cal H}}
\renewcommand{\theequation}{{\thesection}.{\arabic{equation}}}
\def\la{\lambda}
\def\ka{\kappa}
\def\vac{|\mbox{vac}\rangle}
\def\lvac{\langle\mbox{vac}|}
\def\intt{\int_{-\infty}^\infty}
\def\h{\hbar}
\def\stackupb#1#2#3{\ \mathrel{\mathop{#1}\limits_{#2}^{#3}}}
\def\Ga#1{\Gamma\left(#1\right)}
\def\dif#1#2{{\partial #1\over\partial #2}}
\def\psio{\overline\psi}
\def\b{\theta}
\def\a{\alpha}
\def\al{\alpha}
\def\spin#1#2{\left(\begin{array}{c}#1\\#2\end{array}\right)}
\def\te{\theta}
\def\d{\delta}
\def\g{\gamma}
\def\spin#1#2{\left(\begin{array}{c}#1\\#2\end{array}\right)}
\def\into{\int_{0}^\infty}
\def\h{{1\over 2}}
\def\wZ{\widehat{Z}}
\let\dag=\dagger
\def\Z{{\cal Z}}
\def\wwZ{\widehat{\cal Z}}
\def\pnu{\nu'}
\def\bh{\beta}
\def\T{{\cal T}}
\let\appox=\sim
\def\vp{v}
\def\qp{q}
\def\lp{l}
\def\Rt{\tilde{R}}
\def\qt{\tilde{q}}
\def\Tt{\tilde{\T}}
\def\LL{{\cal L}}
\def\pss{\overline{\Psi}}
\def\df#1{\partial_{ #1}}
\def\di#1{\partial^{ #1}}
\def\vacph{|\mbox{vac}\rangle_{\rm ph}}
\def\lvacph{_{\rm ph}\langle\mbox{vac}|}
\def\Tr{{\rm Tr}\,}
\def\AA{{\cal A}(\widehat{sl}_2)}
\def\aa{\alpha}
\def\Ad{{\rm Ad}}
\def\End{{\rm End}}
\def\tT{t}
\def\Rr{R}
\def\R{\CR}
\def\bvac#1{|#1\rangle_{b}}
\def\Q{{{\cal Q}}}
\def\P{{{\cal P}}}
\def\H{\CH_R}
\let\tih=\pi
\let\Ael=\AA
\def\rav#1{\stackrel{\mbox{\tiny {\rm \r{#1}}}}{=}}
%
%
%
\def\vpint{\int_{-\infty}^\infty\hspace{-20.5pt}-\hspace{11.5pt}}
\def\vpinto{\int_{0}^\infty\hspace{-18.6pt}-\hspace{9.6pt}}
%

\begin{center}
\hfill ITEP-TH-75/98, CLNS 99/1607\\
\hfill hep-th/9904082\\

\bigskip\bigskip
{\Large\bf Angular Quantization of the Sine-Gordon Model\\
at the Free Fermion Point}\\
\bigskip
\bigskip
{\bf
S. Khoroshkin$^{\star}$\footnote{E-mail: khoroshkin@heron.itep.ru},\
A. LeClair$^{*}$\footnote{E-mail: leclair@mail.lns.cornell.edu},\
S. Pakuliak$^{\star\bullet\circ}$}\footnote{E-mail: pakuliak@thsun1.jinr.ru}\\
\bigskip
$^\star$
{\it Institute of Theoretical \& Experimental Physics, 117259 Moscow,
Russia}\\
$^{*}$
{\it Newman Laboratory, Cornell University, Ithaca, NY 14853-5001, USA}\\
$^\bullet$
{\it Bogoliubov Laboratory of Theoretical Physics, JINR,
141980 Dubna, Moscow region, Russia}\\
$^\circ$
{\it Bogoliubov Institute for Theoretical Physics, 252143 Kiev, Ukraine}
\bigskip
\bigskip
\end{center}
\begin{abstract}
The goal of this paper is to analyse the method of angular quantization for
the  Sine-Gordon model at the free fermion point, which is one of the most
investigated models of the two-dimensional integrable field theories. The
angular quantization method (see  hep-th/9707091) is a continuous analog of
the Baxter's corner transfer matrix method. Investigating the canonical
quantization of the free massive Dirac fermions in one Rindler wedge we
identify this quantization with a representation of the infinite-dimensional
algebra introduced in the paper q-alg/9702002 and  specialized to the free
fermion point. We construct further the main ingredients of the SG theory in
terms of the representation theory of this algebra following the approach by
M.Jimbo, T.Miwa et al.
\end{abstract}

\footnotesize
\tableofcontents
\normalsize

\newpage

\setcounter{footnote}{0}
\setcounter{equation}{0}
\section{Introduction}

The Sine-Gordon (SG) model in two-dimensional Minkowski space-time is
described by the action\footnote{
We  have rescaled $\beta \to \sqrt{4\pi} \beta $ in comparison
with the usual convention, so that the free fermion (FF) point occurs
at $\bh^2 = 1$.}
\beq\label{SinGor}
S_{\rm SG}=\frac{1}{4\pi} \int dt\,dx\ \sk{
\frac{1}{2}
\frac{\partial^2\Phi(x,t)}
{\partial t^2}
-\frac{1}{2}
\frac{\partial^2\Phi(x,t)}
{\partial x^2}
+\frac{m^2}{\bh^2}\ \sk{\cos(\bh\Phi(x,t))-1}
}.
\eeq
The quantum SG theory is perhaps the most fundamental of the integrable
quantum field theories in two dimensions,
and thus plays an important role in the development
of new methods.
The  $S$-matrix of
soliton-antisoliton scattering was obtained
in \cite{ZZ}.
This $S$-matrix (see \r{S-mat})
depends on so called renormalized coupling constant $\xi$
and the relation of this parameter to  the SG coupling constant $\bh$ is
\beq\label{ren-coup}
\xi=\frac{\bh^2}{2-\bh^2}\ .
\eeq
The quantum SG model is a superrenormalizable theory for the real values
of the coupling constant $0<\bh^2<2$ which corresponds to the restriction
to the real positive values of $\xi$, $0<\xi<\infty$.
The regime $1<\xi<\infty$ is the  breatherless one, where solitons and
antisolitons do not form bound states.
One can see that modulo the overall  scalar factor and for appropriate choice
of the multiplicative spectral parameter $z=e^{-\theta/\xi}$,
where $\theta$ is a rapidity of the particles,
the soliton-antisoliton $S$-matrix can be
written in the form
\beq\label{s-mul}
S(\theta,\xi)=\rho(\theta,\xi)
\left(\begin{array}{cccc}
{zq-z^{-1}q^{-1}}&0&0&0\\
0&{z-z^{-1}}&{q-q^{-1}}&0\\
0&{q-q^{-1}}&{z-z^{-1}}&0\\
0&0&0&{zq-z^{-1}q^{-1}}
\end{array}\right),\quad q=\exp\sk{\pi i\frac{\xi+1}{\xi}}
\eeq
which signifies in particular a quantum group symmetry of the
Hilbert space of states of the model with respect to the
finite dimensional quantum group $U_q(sl_2)$ \cite{RS,AL}.

The SG model was also one of the first continuous integrable models where the
quantum
inverse scattering method (QISM) was tested. It was shown in
the paper \cite{FST} that the quantum monodromy
matrices $\T(u)$  satisfy the commutation relation
\beq\label{YBt}
R(u_1-u_2,\xi)\T_1(u_1)\T_2(u_2)=\T_2(u_2)\T_1(u_1)R(u_1-u_2,\xi)\ ,
\eeq
where the  $R$-matrix has the same structure as in \r{s-mul}
(see \r{RR-mat} for the exact formula)
 in terms  of
additive an spectral parameter $u$ $\sk{z=e^{-u/(\xi+1)}}$,  but with
deformation parameter replaced by
\beq\label{q'-def}
q'=\exp\sk{\pi i\frac{\xi}{\xi+1}}.
\eeq
The equation \r{YBt} implies that
\beq\label{trace}
[\tr\,\T(u_1),\tr\,\T(u_2)]=0
\eeq
and signifies that after proper expansion of the quantity $\tr\,\T(u)$
with respect to the spectral parameter $u$ it generates the local
integrals of motion and \r{trace} shows that they are in involution.
Note that even  at the FF point where $\bh^2=1$ the
$R$-matrix in the commutation relation of monodromy matrices \r{YBt}
is nontrivial
since $q'=i$; this can be traced to the fact that the monodromy matrix
is constructed from the fields $\exp(i\Phi /2)$, which are non-local
in terms of the fermions since the fermion bilinear is $\exp(i\Phi )$.

As we see, the SG model naturally contains two quantum group symmetries,
with different deformation parameters related by the duality transformation
\r{dual-tr}. An attempt to explain this phenomena was made in
\cite{Lu1} in the framework of the bosonization technique in massive
integrable field theories. This approach was generalized then for the
lattice integrable models \cite{FIJKMY}. Following the ideas presented
in these papers a screening current algebra was proposed in \cite{KLP}.
The specific coalgebraic properties of this infinite-dimensional algebra
allowed to reconstruct the bosonization approach of \cite{Lu1} from
algebraical analysis of the representation theory of the screening current
algebra.

Essential progress toward  understanding  quantum integrable models
in the  infinite volume limit was made in the framework of Baxter's corner
transfer matrix (CTM) method  \cite{B}. It was observed that the CTM
of some lattice integrable models
in the
infinite volume limit has equidistant spectrum bounded from below
and so can be described
by the infinite set of oscillators. This fact allows one to develop a new
approach to quantum integrable models on the lattice. This was done
by the Kyoto group for the XXZ
model in the anti-ferroelectric regime \cite{JM}.
The model was completely solved, namely, the correlation
functions of local operators and form-factors of local operators
were calculated explicitly, using infinite-dimensional representations
of quantum affine algebra $U_{q}(\widehat{sl}_2)$ with real parameter
of deformation satisfying $-1<q<0$.
One of the main ideas
of the construction is to divide the total Hilbert space of the model,
which is identified in the infinite volume  limit with an infinite product
of two-dimensional spaces where local operators act,
\beq\label{hilXXZ}
\CH_{\rm XXZ}\approx
\cdots\CC^2\ot\CC^2\ot\CC^2\ot\CC^2\ot\CC^2\ot\CC^2\cdots
\eeq
 into two
semi-infinite products of these spaces
\beq\label{hilXXZ1}
\CH_{\rm XXZ}\approx\bigr(\cdots\CC^2\ot\CC^2\ot\CC^2\bigl)\ot
\bigr(\CC^2\ot\CC^2\ot\CC^2\cdots\bigl)\approx
\CH^*_{\rm CTM}\ot\CH_{\rm CTM}=\End\sk{\CH_{\rm CTM}}
\eeq
which are denoted by $\CH_{\rm CTM}$ and where
the corner transfer matrix
 acts naturally.
 Each of these semi-infinite products
is identified with level 1 and level $-1$ integrable modules of
 $U_{q}(\widehat{sl}_2)$, where operators of the algebra act naturally.
The decomposition \r{hilXXZ1} led in particular to the identification of the
states in the Hilbert space $\CH_{XXZ}$ with the operators
acting in $\CH_{CTM}$.
The space $\End\sk{\CH_{\rm CTM}}$ is equipped with a natural scalar product
$(A,B)=\Tr_{\CH_{\rm CTM}}AB$ and the vacuum vector in
$\CH_{\rm XXZ}$ is defined as $(-q)^
{H_{\rm CTM}}$, where
$H_{\rm CTM}$ is a corner transfer matrix hamiltonian.

The representation theory of the quantum affine algebra
$U_{q}(\widehat{sl}_2)$ provides certain
operators which intertwine its action in $\CH_{\rm CTM}$ (type I and type II
intertwining operators). Type II intertwining operators are used for the
construction of the basis of asymptotic states in
$\End\sk{\CH_{\rm CTM}}$, and type I operators are used for the
construction of the
transfer matrix and the local hamiltonian in this picture.
Moreover, the adjoint action of the elements of the quantum affine
algebra in $\End\sk{\CH_{\rm CTM}}$ describe a level 0
$U_{q}(\widehat{sl}_2)$ symmetry of the model.
As a consequence, the form-factors of the local operators and
correlation functions of their product are presented in a form of
certain multiple integrals, which come as a trace over
$\CH_{\rm CTM}$ of certain products of the intertwining
operators.

In  the continuous
integrable models an approach to implement Baxter's CTM method was
developed in the papers \cite{Lu1,BrL}
and was based on
 the method of the angular quantization. The total Hilbert space
of the continuous quantum integrable model in infinite volume
was supposed to be embedded into a tensor product
\beq\label{decomp}
\CH \hookrightarrow \CH_L \otimes \CH_R\ ,
\eeq
where $\CH_L$ ($\CH_R$) are the Hilbert spaces of the
quantization in the left (right) wedge.
The right Rindler wedge (RRW) in two-dimensional Minkowski space-time is
\beq\label{rrw}
(x^0)^2-(x^1)^2<0,\quad x^1 >0\ ,
\eeq
where $x^0$ is a time and $x^1$ is coordinate,
while the left Rindler wedge (LRW)
\beq\label{lrw}
(x^0)^2-(x^1)^2<0,\quad x^1 <0\ .
\eeq
Let us fix the parametrization of space-time coordinates in RRW
\beq\label{para}
x^0=r\,\sh\,\a, \quad x^1=r\,\ch\,\a,\quad r\geq0,\quad \a\in\RR\  .
\eeq
With this parametrization, the coordinates $x^0, x^1$ cover
the RRW, since $x^1 >0$.
The LRW
is formally
obtained by the rotation $\alpha \to \alpha -i\pi$ or by applying
the operator $e^{\pi  K}$
where $K$ is Lorentz boost generator $K = -i \partial_\alpha$.
The space $\CH_L$ can be identified with the dual to $\CH_R$
and so the states in the total Hilbert space can be realized as
the operators in $\CH_R$.

It was suggested
in \cite{Lu1}
to realize $\CH_R$ for the
SG model as a Fock space with a natural action of the operators satisfying
the commutation relations of
Zamolodchikov-Faddeev (ZF) algebra. Further, in \cite{KLP} these operators
were identified with intertwining operators of the scaled elliptic
algebra $\AA$ which can be observed in the bosonization picture
\cite{Lu1} by the presentation using screening currents.
One of the main arguments in favor of these mathematical constructions was the
coincidence of form factors of certain local operators in SG theory with
trace calculations in $\CH_R$.

In this paper, we try to develop the method of the angular
quantization in two directions.
First, we analyze the SG model in RRW at the free fermion point, where the
canonical quantization can be done explicitly. We see here that the usual
conserved charges \cite{Le} diverge and the only chance to get a rich algebra
of symmetries is to use a certain
analytical continuation of the conserved charges, or
equivalently, the scattering data. In this case the bosonization \cite{Lu1}
naturally appears. We see further that in order to close the algebra, we are
forced to use the currents with dual monodromy properties and the algebra of
(nonlocal) conserved currents which we find here coincide with
specialization of the scaling elliptic algebra $\AA$ proposed in \cite{KLP}
and specialized to the free fermion point ($\xi=1$).

Second, we go into further details of the description of the continuous SG
model analogous to the group-theoretical description of the space of states
in the  XXZ model \cite{JM}. We show that starting from level one
representation of the scaling elliptic algebra $\AA$ we can correctly define
the vacuum, the asymptotic states and operators which act on the space
of the asymptotic states, namely,  the transfer matrix, the hamiltonian,
the local integrals of motion. Contrary to the lattice case they are given
now via coefficients of the asymptotic expansion of the family of
commuting operators. We define the adjoint action of the algebra $\AA$ on
the space of states and show that known symmetries of this space related
to the conserved nonlocal currents \cite{L,BL}
and formulated in terms of quantum affine algebras at level zero,  can be
obtained from this adjoint action by the asymptotical expansion.
 Let us roughly explain  this description.

The total Hilbert space $\CH$ is supposed to be divided as in \r{decomp}.
The spaces $\CH_R$ and $\CH_L$ are level 1 and level $-1$ highest weight
modules over the algebra $\AA$ so the states in $\CH$ can be identified with some
operators in $\CH_R$. In particular, the physical vacuum state $\vacph$
is identified with boost operator
\beq\label{vacph}
\vacph=e^{\pi K}=e^{-i\pi\partial_\al},
\eeq
where $\al$ is angular time in RRW
and the states $|\theta_1,\ldots,\theta_n\rangle_{\ep_1,\ldots,\ep_n}$
are identified with the product
\beq\label{stat}
|\theta_1,\ldots,\theta_n\rangle_{\ep_1,\ldots,\ep_n}=
Z^*_{\ep_1}(\theta_1)\ldots Z^*_{\ep_n}(\theta_n)e^{\pi K}\ ,
\eeq
where $Z^*_{\ep}(\theta)$ are certain `twisted' intertwining operators of the
screening currents algebra $\AA$, which also act in $\CH_R$.
The adjoint action of the algebra $\AA$ is not standard because
this algebra is not a Hopf algebra. Indeed, the
commutation and comultiplication
relations of the algebra $\AA$ in terms of $L$-operators
can be written in the form
\beq\label{scr-com}
\R(u_1-u_2,\xi+c)L_1(u_1,\xi)L_2(u_2,\xi)=L_2(u_2,\xi)L_1(u_1,\xi)
\R(u_1-u_2,\xi)
\eeq
\beq\label{scr-coal}
\Delta^{\rm op} L(u,\xi)=L(u- i\pi c^{(2)}/4,\xi+c^{(2)})
\ {\dot\otimes}\ L(u+ i\pi c^{(1)}/4,\xi)
\nn
\eeq
where $\R(u,\xi)$ means $R$-matrix defined by \r{R-mat}
and $c$ is a central element of the algebra $\AA$.
Note that $R$-matrices in the left and right hand sides of \r{scr-com}
differ by the central element of the algebra, which signifies that the
algebra under consideration is not coassociative.
This algebra is not a usual Hopf algebra. Nevertheless, a coalgebraic
structure  of this algebra was used in \cite{KLP} to construct
the intertwining operators for  highest weight
modules over this algebra at the value of the central element
$c=1$. There are also some indications that
this screening current algebra is a quasi-Hopf algebra \cite{Drquasi}
(see
papers \cite{JKOS1-2} on the lattice variants of this algebra).

The adjoint action has the different form on the subspaces $\CH_i\in\CH$,
$i=0,1$ of even and odd number of particles and includes the involution of
the algebra $\AA$
\beq
\iota \sk{L(u)}=
\sigma_z   L(u)    \sigma_z\ .
\label{L-oper-rel}
\eeq
For the state $X_i\in\CH_i$, $i=0,1$ it is defined as follows
\beq
\Ad_{L(u;\xi)}\cdot X_{k}=
 \iota\sk{L^{-1}(u+i\pi c/4;\xi)}\,X_{k}\,\iota^{k+1}
\sk{L(u-i\pi +i\pi c/4;\xi)}\ .
\label{adj}
\eeq
We prove that so defined adjoint action realizes the level zero
representation of the algebra $\AA$ onto the space of states $\CH$,
such that $n$-particle states compose $n$-fold tensor products of
two-dimensional representations.
The quantum affine symmetry of the Hilbert space $\CH$ found in \cite{BL}
can be realized via the asymptotical expansion of the adjoint action of the
currents of the algebra $\AA$.

The paper is organized as follows. In the second section we consider
the canonical quantization of the
SG model at FF point and its specialization to
right Rindler wedge. Then we construct the nonlocal integrals of motion
and develop bosonization of all the objects in terms of these
integrals of motion. The third section is devoted to the description of the
screening currents
algebra for general value of SG coupling constant satisfying $1<\bh<\sqrt{2}$.
In the last section we develop the angular quantization scheme
in full aspect; for example, we construct the monodromy matrix on the
total Hilbert space and
investigate some of its properties.

\setcounter{equation}{0}
\section{Canonical quantization}

\subsection{Sine-Gordon model at free fermion point}

It is well known \cite{Col} that the SG model
 model with the action \r{SinGor} is equivalent on the
quantum level to the massive Thirring model defined by the action
\beq\label{Thir}
S_{\rm Th}=\int dx^0dx^1\
\left[\h\left(\pss(x) i\g^\mu\df \mu \Psi(x)-
\df \mu\pss(x) i\g^\mu \Psi(x)\right)-m\pss(x)\Psi(x)
-\frac{g}{2}\sk{\pss(x)\g^\mu\Psi(x)}^2
\right]
\eeq
where $g=\frac{\pi(1-\xi)}{2\xi}$.
The equivalence is established by the following bosonization rules
\beq\label{S-T}
\frac{\bh}{2\pi}\varepsilon^{\mu\nu}\partial_{\nu}\Phi(x^0,x^1)=
\pss(x^0,x^1)\g^\mu\Psi(x^0,x^1),\quad
\cos(i\bh\Phi(x^0,x^1))= \pss(x^0,x^1)\Psi(x^0,x^1)
\eeq
where $\varepsilon^{\mu\nu}$ is antisymmetric tensor normalized
$\varepsilon^{01}=1$.

At the FF point ($\xi=1$) the interaction in the Thirring model vanishes
and its lagrangian becomes a lagrangian of free massive Dirac
fermions
\beq
\LL(x^0,x^1)=
\left[\h\left(\pss(x) i\g^\mu\df \mu \Psi(x)-
\df \mu\pss(x) i\g^\mu \Psi(x)\right)-m\pss(x)\Psi(x)\right],
\eeq
where $\pss(x)=\Psi^\dag(x)\g^0$ is a Dirac conjugated spinor.
We fix the $\g$-matrices to be
$$
\g^0=\left(\begin{array}{cc} 0& -i\\ i&0\end{array}\right),
\quad
\g^1=\left(\begin{array}{cc} 0& i\\ i&0\end{array}\right).
$$

The complete set of  solutions to the corresponding linear
equation of motion can be chosen in the form
\beq
\Psi_\b(x^0,x^1)=\sqrt{m\over 2}\spin {e^{\b/ 2}}{ie^{-\b/2}}
e^{-im\ch(\b)x^0+im\sh(\b)x^1}
\label{ps}
\eeq
and satisfies the completeness relation with respect
to the scalar product
\beq
(\Psi_{\b},\Psi_{\b^\prime})=
\intt dx^1 \pss_\b(x^0,x^1)\g^0\Psi_{\b^\prime}(x^0,x^1)=\d(\b-\b^\prime).
\label{nops}
\eeq
Note that the solutions \r{ps} are anti-periodic with respect to
the shift $\b\to\b+2\pi i$.

The completeness relation \r{nops}  allows one  to quantize the Dirac field
$\Psi(x)\equiv\Psi_+(x)$ and its hermitian conjugate
$\Psi^\dag_+(x)\equiv\Psi_-(x)$
\bea
\Psi_+(x^0,x^1)=\intt d\b\left[
c(\b)\Psi_\b(x)
+d^{\dag}(\b)\Psi^*_\b(x)\right],
\label{ts1+}
\eea
\bea
\Psi_-(x^0,x^1)=\intt d\b\left[
d(\b)\Psi_\b(x)
+c^{\dag}(\b)\Psi^*_\b(x)\right],
\label{ts1-}
\eea
by imposing equal time
anticommutation relations
\beq\label{commm}
\{\psi_+(x^0,x),\psi_-(x^0,x')\}=2\pi\d(x-x'),\quad
\{\psio_+(x^0,x),\psio_-(x^0,x')\}=2\pi\d(x-x'),
\eeq
where $\psi_\pm(x^0,x^1)$ and $\psio_\pm(x^0,x^1)$ are components
of spinor $\Psi_\pm(x^0,x^1)$.
One  can verify now that \r{commm}
and the normalization
condition \r{nops}
imply the standard
anticommutation relations
\beq
\{c(\b),c^\dag(\b')\}=\d(\b-\b'), \quad
\{d(\b),d^\dag(\b')\}=\d(\b-\b')
\label{A+A-}
\eeq
and all others are trivial.

The Hilbert space $\CH$ of this model is defined by application of the
creation operators $c^\dag(\b)$ and $d^\dag(\b)$ to the `physical'
vacuum vector $\vacph$ annihilated by the operators $c(\b)$ and
$d(\b)$.

The integrals of motion are constructed from the conserved
currents:
$$
\dif{J_y}{\bar y}-\dif{J_{\bar y}}{y}=0
$$
and are given by the contour integral
\beq\label{charges}
Q^J=\int \sk{dy\ J_y + d\bar y\ J_{\bar y}}\ .
\eeq
The charge $Q^J$ is conserved along the evolution which is
`orthogonal' to the contour in the definition of $Q^J$.
In the standard quantization picture this contour is chosen to be
equal time line $x^0=\hbox{const}$  in the space-time.
The charges which are conserved along the evolution with respect to the
time $x^0$ can be obtained from the currents
\bea
 J_y^\pm&=&(D \psi_\pm) \psi_\pm,\quad
 J_{\bar y}^\pm=(D\psio_\pm) \psio_\pm,\label{iii1}\\
 J_y^0&=&(D\psi_- )\psi_+,\quad
 J_{\bar y}^0=(D\psio_- ) \psio_+,\label{iii2}
\eea
or equivalently from
\bea
\tilde J_y^\pm&=&\psi_\pm (D \psi_\pm),\quad
\tilde J_{\bar y}^\pm=\psio_\pm (D \psio_\pm),\label{ii1}\\
\tilde J_y^0&=&\psi_- (D \psi_+),\quad
\tilde J_{\bar y}^0=\psio_- (D \psio_+),\label{ii2}
\eea
for the operator $D=\partial_y^n$ or $\partial_{\bar y}^n$,
$n\geq0$.
Let us denote the neutral charges which correspond to the first operator
as $I_n$ and to the second as $\bar I_n$. The charged conserved
quantities we denote by $I^\pm_n$ and $\bar I^\pm_n$ respectively.
They have the explicit expressions in terms of the
operators acting in total Hilbert space:
\beq\label{int-n}
I_n=m^n\intt d\theta\ e^{n\theta}\sk{c^\dag(\theta)c(\theta)+
d^\dag(\theta)d(\theta)},\quad
\bar I_n=m^n\intt d\theta\ e^{-n\theta}\sk{c^\dag(\theta)c(\theta)+
d^\dag(\theta)d(\theta)}
\eeq
and similar formulas for $I^\pm_n$ and $\bar I^\pm_n$.
The Hamiltonian $H$ which describes the evolution of the quantum
fields \r{ts1+} and \r{ts1-} with respect to the time $x^0$ is given
by the sum $(I_1+\bar I_1)/2$ and has eigenvalue $m\ch\,\theta$ on the
one-particle states generated by $c^\dag(\theta)$ and $d^\dag(\theta)$
from physical vacuum $\vacph$.

However there is no a direct way to quantize the SG field $\Phi(x^0,x^1)$
at the FF point
using the quantization of the Dirac fermion fields $\Psi(x^0,x^1)$.
In particular, it is difficult to construct the realization of the
commutation relations \r{YBt} directly in the infinite volume limit
using the canonical anticommutation relations \r{commm}  and
without referring to the lattice regularization.

On the other hand the canonical quantization of the free massive Dirac
fermions in RRW allows one to construct the operators which are
building blocks of the
angular quantization method. This will be done in the next subsections
with the main goal being to demonstrate the  nonabelian
symmetry algebra which appears in the angular quantization approach
to the SG model.

\subsection{Free fermions in Rindler wedge}

Let us solve the equation of motion for free massive Dirac field
in RRW using the parametrization \r{para}. The solution to the Dirac
equation of motion
normalized with respect to the scalar product
\bea
(\Psi,\Psi')=\into dr\ \sk{e^{-\a}\bar \psi\bar \psi'+e^{\a}
\psi\psi'}\ ,\quad
 \Psi=\sk{\begin{array}{c}\bar \psi\\ \psi \end{array}}
\label{scalar}
\eea
is given
in
terms of MacDonald func\-tions
\beq
\Psi_\nu(r,\a)={\sqrt{m}\over\sqrt{\pi} \Gamma(i\nu+1/2)}
\spin{\ee^{(2i\nu+1/2)\a} K_{i\nu+1/2}(mr)}
{\ee^{(2i\nu-1/2)\a} K_{i\nu-1/2}(mr)},\quad \nu\in\RR\ ,
\label{sol}
\eeq
and has exponentially decreasing  asymptotics in RRW when $r\to\infty$.
On the other hand we observe that rotation of angular time
$\al$ by $2\pi i$ which corresponds to the path around origin in
euclidean plane multiplies the solution \r{sol} by the factor
$-e^{-2\pi\nu}$. This signifies the fact
 that the space of functions used in canonical and
angular quantization are completely different.
Nevertheless, the completeness relation
\beq
(\Psi_\nu,\Psi_{\nu'})=\delta(\nu+\nu')
\label{orth}
\eeq
allows one  to quantize the Dirac fields in RRW
\beq
\Psi_\pm(r,\a)=\spin{\bar\psi_\pm(r,\a)}{\psi_\pm(r,\a)}=\intt d\nu\
b_\pm(\nu)\Psi_\nu(r,\a)
\label{rw+}
\eeq
by imposing the equal `time' ($\al=\hbox{const}$) anticommutation
relations
\beq
\{\psi_+(r,\a),\psi_-(r',\a)\}=-\ee^{-\a}\d(r-r'),\quad
\{\psio_+(r,\a),\psio_-(r',\a)\}=-\ee^{\a}\d(r-r')
\label{coa}
\eeq
which are equivalent to
\beq
\{b_\pm(\nu),b_\pm(\nu')\}=0,\quad
\{b_+(\nu),b_-(\nu')\}=\delta(\nu+\nu')\ .
\label{stand}
\eeq

Rindler fermionic Fock space $\CH_R^f$
is defined by the vacuum state $\vac_f$ which
satisfies
\beq\label{Fock}
b_\pm(\nu)\vac_f=0,\quad \nu>0\ .
\eeq
The left vacuum vector $_f\lvac$ is correspondingly defined:
\beq\label{lFock}
_f\lvac b_\pm(\nu)=0,\quad \nu<0\ .
\eeq

\subsection{Scattering transform}

For the quantum fermionic fields \r{rw+} we introduce the
scattering transform \cite{Lu-pri}
\beq\label{sc-tr}
\Psi(r,\al)\to\Lambda_\pm(\theta,\al)= \frac{\sqrt{m}}{2\sqrt\pi}\ \into dr\
e^{-mr\ch\theta}\sk{
\overline\psi_\pm(r,\al)e^{(\theta-\al)/2}+\psi_\pm(r,\al)e^{(\al-\theta)/2}}
\ ,
\eeq
where $\theta\in\CC$ is spectral parameter. Using the free fermion
equation of motion in RRW we can verify that the dependence of the operators
$\Lambda_\pm$ on the angular time $\al$ reduces to a simple shift of the
spectral parameter
\beq\label{stpr}
\Lambda_\pm(\theta,\al)=\Lambda_\pm(\theta+\al)\ ,
\eeq
where
$\Lambda_\pm(\theta)$ is the value of the scattering transform at the
initial time, say $\al=0$. It is clear that the scattering transform \r{sc-tr}
is not defined  for all values of the spectral parameter $\theta$. For
example, if the solutions $\overline\psi_\pm(r,\al)$ and $\psi_\pm(r,\al)$
have the constant asymptotics when $r\to\infty$ then the integral in
\r{sc-tr} is convergent if $|\Im\,\theta|<\pi/2$, which follows from the
inequality $\Re\,\ch\,\theta>0$. However, the solutions \r{sol} of the Dirac
equation in RRW have exponentially decreasing asymptotics. Using the fact
that the leading term of the asymptotic of the MacDonald function $K_x(z)$
when $z\to\infty$ does not depend on the index $x$ and is proportional to
$z^{-1/2}e^{-z}$ we find that the inequality mentioned above is replaced by
the more weak inequality
\beq\label{ineq} \Re\,\ch\,\theta>-1\ . \eeq
The solution
of \r{ineq} defines a larger
 domain of existence of the scattering transform than
specified above, namely
\beq\label{dom1}
|\Im\,\theta|<\pi/2+\epsilon,\quad\hbox{where}\quad
\epsilon=\pi/2-\arccos\sk{\sk{\ch\,\Re\,\theta}^{-1}}
\eeq
so the domain of the
possible values of the spectral parameter is a strip whose width depends
on the value of $\Re\,\theta$.

An important consequence of this observation is the fact that the points
$\Im\,\theta=\pm\pi/2$ are {\it always} in the domain of existence of the
scattering transform. This leads to the fact that the vacuum expectation
value of the product $\Lambda_\pm(\theta) \Lambda_\mp(\theta')$ is
a well defined meromorphic function
  in the domain $|\Im\,(\theta-\theta')|\leq \pi+\ep$ for
some positive number $\ep$. Using the expression
of the scattering data operators $\Lambda_\pm(\theta)$ in terms of the
fermionic operators $b_\pm(\nu)$
\beq\label{zz5}
\Lambda_\pm(\theta)=
\frac{1}{2\pi}\intt d\nu\ b_\pm(\nu)\Gamma\sk{\frac{1}{2}
-i\nu}e^{i\nu\theta}
\eeq
we can calculate this function explicitly:
\beq\label{vev-sc}
\langle\Lambda_\pm(\theta)\Lambda_\mp(\theta')\rangle=
\frac{1}{4\pi^2}\beta\sk{\frac{\theta-\theta'+\pi i}{2\pi i}},
\eeq
where the $\beta$-function is $\beta(x)=
\partial_x\ln\sk{\left.\Gamma\sk{\frac{x+1}{2}}
\right/\Gamma\sk{\frac{x}{2}}}$.
Since the function \r{vev-sc} has the poles only in the points
$\theta=\theta'-\pi i(2k+1)$, $k=0,1,\ldots$ the domain of the `existence'
of this function can be extended to $\Im(\theta-\theta')>-3\pi$ with a simple
pole at the point $\theta=\theta'-\pi i$. The function
$\langle\Lambda_\pm(\theta)\Lambda_\mp(\theta')\rangle$ is given by the
meromorphic function \r{vev-sc} in this domain.
An
immediate consequence of this fact is the anticommutation relation
\beq\label{alLam}
\{\Lambda_+(\theta),\Lambda_-(\theta')\}=\frac{\pi}{\ch((\theta-\theta')/2)},
\quad |\Im(\theta-\theta')|<3\pi\ .
\eeq

The scattering transform \r{sc-tr} describes an evolution of the initial data
(the quantum fields $\Psi_\pm(r,0)$ at initial value of the angular time
$\al=0$) with respect to this angular time.
 Since this  evolution reduces to a simple shift in the spectral parameter
one can easily restore
the quantum fields $\Psi_\pm(r,\al)$ at arbitrary time $\al$
by solving inverse scattering problem, restoring quantum fields
$\Psi_\pm(r,\al)$ from the operators $\Lambda_\pm(\theta)$.

This can be done using the operators $\Z_\pm(\theta)$ related
to the operators $\Lambda_\pm(\theta)$ by the integral transform
\beq\label{ZL7}
\Lambda_\pm(\theta)=\frac{1}{4\pi}\intt d\theta'\ \frac{\Z_\pm(\theta')}
{\ch((\theta-\theta')/2)}
\eeq
which can be inverted as follows
\beq\label{LZ7}
\Z_\pm(\theta)=\Lambda(\theta+\pi i)+\Lambda(\theta-\pi i)\ .
\eeq
The inverse scattering problem has a solution in terms of the operators
$\Z_\pm(\theta)$:
\beq\label{psi-Z}
\Psi_\pm(r,\al)={\sqrt{m}\over2\sqrt{\pi}}\intt d\theta\
\Z_\pm(\theta+\al)e^{-mr\ch\,\theta}
\spin{e^{(\theta+\al)/2}}{e^{-(\theta+\al)/2}}.
\eeq
One can observe that substitution of \r{psi-Z} into \r{sc-tr} leads to the
integral transform \r{ZL7}.

Formula \r{LZ7} allows one to calculate the vacuum expectation values
of the different operators
\bea
\langle \Lambda_\pm(\theta_1)
\Z_\mp(\theta_2)\rangle &=&
\langle \Z_\pm(\theta_1)
\Lambda_\mp(\theta_2)\rangle=
-\frac{1}{2\pi i}\
\frac{1}{\theta_1-\theta_2}, \quad \Im(\theta_1-\theta_2)>-2\pi\ ,\nn\\
\langle \Z_\pm(\theta_1)
\Z_\mp(\theta_2)\rangle &=& -\frac{1}{2\pi i} \left[
\frac{1}{\theta_1-\theta_2-i\pi}+\frac{1}{\theta_1-\theta_2+i\pi}
\right],\quad \Im(\theta_1-\theta_2)>-\pi\ .
\label{vev}
\eea
These formulas allow to verify that the canonical anticommutation relations
\r{coa} follow from the solution of the inverse scattering
problem \r{psi-Z}. The formula \r{vev} demonstrates also that
the operators $\Z_\pm(\theta)$ anticommute for real values of the spectral
parameter $\theta$.

We would like to remark here that the operators
$\Z_\pm(\theta)$ being expressed in terms of the fermionic operators
$b_\pm(\nu)$
\beq\label{ZZ6}
\Z_\pm(\theta)=
\intt d\nu\ \frac{b_\pm(\nu)}{\Gamma\sk{\frac{1}{2}
+i\nu}}e^{i\nu\theta}
\eeq
should not be understood literally, but rather as a certain normal ordering
expression, where  the normal ordering is dictated by the prescription
\r{vev} on the domains of analyticity of the products of the scattering data
operators.
The naive use of the vacuum expectation
value
$$
\langle b_\pm(\nu)b_\mp(\nu')\rangle=\delta(\nu+\nu')\Theta(\nu)
$$
where $\Theta(\nu)$ is the step-function,
in order to calculate \r{vev}
does not allow to find the domain where the vacuum
expectation value $\langle \Z_\pm(\theta_1)
\Z_\mp(\theta_2)\rangle$ is defined since this information
is encoded in the analytical properties of the
scattering transform.

In order to obtain \r{coa} from \r{vev} it is convenient to introduce
the operators
$Z_\pm^*(\theta)$ and $Z_\pm(\theta)$ as a shift
by $\pm\pi i/2$ of the operators $\Z_\pm(\theta)$ using the freedom
to move the contour in the integral representation \r{psi-Z}:
\beq\label{ZF-oper}
Z^*_\pm(\theta)=\Z_\pm(\theta-\pi i/2),\quad
Z_\pm(\theta)=\Z_\mp(\theta+\pi i/2)\ .
\eeq
Then the fermion fields \r{psi-Z} can be rewritten in the two equivalent
form:
\bea
\Psi_\pm(r,\al)&=&{\sqrt{m}\over2\sqrt{\pi}}\intt d\theta\
Z_\mp(\theta+\al)e^{-imr\sh(\theta)}
\spin{e^{\pi i/4}e^{(\theta+\al)/2}}{e^{-\pi i/4}e^{-(\theta+\al)/2}}\nn\\
&=&{\sqrt{m}\over2\sqrt{\pi}}\intt d\theta\
Z^*_\pm(\theta+\al)e^{imr\sh(\theta)}
\spin{e^{-\pi i/4}e^{(\theta+\al)/2}}{e^{\pi i/4}e^{-(\theta+\al)/2}}.\nn
\eea
The operators $Z_\pm(\theta)$ or  $Z^*_\pm(\theta)$
acting in angular fermionic or bosonic Hilbert spaces
can be associated with the states in total Hilbert space $\CH$ of the
model. In the next subsections we will identify them with intertwining
and dual intertwining operators for the screening current algebra.
The fact that the pole at the point $\theta_1=\theta_2+i\pi$ does not
produce the restriction on the domain of the analyticity can be seen in
the general situation, because the origin of this pole is the pinching of
the contour in the integral representation of the function
$\langle \Z_\pm(\theta_1)
\Z_\mp(\theta_2)\rangle$ when  $\theta_1\to\theta_2+i\pi$.

\subsection{Integrals of motion}

The operator which describes the evolution with respect
to the angular time $\al$ is the Lorentz boost operator $K$.
In terms of the fields:
$$
K= \frac{i}{2}\into dr\ r\
\Bigl[e^{-\al}
\sk{\psio_-\sk{\partial_r\psio_+}-\sk{\partial_r\psio_-}\psio_+}- e^{\al}
\sk{\psi_-\sk{\partial_r\psi_+}-\sk{\partial_r\psi_-}\psi_+}
+2m \psio_-\psi_+ -2m \psi_-\psio_+
\Bigr]\ .
$$
Using canonical anticommutation relations
\r{coa} one can find the action of this operator on the components
of the Dirac spinor:
\bea
{[}K, \psio_\pm(r,\al){]}&=&
i\frac{\partial\psio_\pm(r,\al)}{\partial\al}-
\frac{i}{2}\psio_\pm(r,\al) ,\nn\\
{[}K, \psi_\pm(r,\al){]}&=&
i\frac{\partial\psi_\pm(r,\al)}{\partial\al}+
\frac{i}{2}\psi_\pm(r,\al)\  .
\label{b-a}
\eea
Note that the boost operator acts differently on the different
components of spinors $\Psi_\pm(r,\al)$.
The formulas \r{b-a} can be rewritten
as follows
\bea
e^{-i\eta K}\psio_\pm(r,\al)e^{i\eta K}&=&
e^{-\eta/2}\psio_\pm(r,\al+\eta),\nn\\
e^{-i\eta K}\psi_\pm(r,\al)e^{i\eta K}&=&
e^{\eta/2}\psi_\pm(r,\al+\eta)\ .\label{b-ae}
\eea
After rotation of the  fermions around the origin by  $\eta=2\pi i$
both of the equations \r{b-ae} become
\beq\label{b-at}
e^{2\pi K}\Psi_\pm(r,\al)e^{-2\pi K}=-\Psi_\pm(r,\al+2\pi i)\ .
\eeq
In terms of Rindler fermions $b_\pm(\nu)$ the boost operator has a form
\beq\label{boost}
K=\intt d\nu\ \nu\ {:}b_-(\nu)b_+(-\nu){:}\equiv
\intt d\nu\ \nu\ {:}b_+(\nu)b_-(-\nu){:}
\eeq
and yields the value of continuous mode $\nu$
\bea
{[}K,b_\pm(\nu){]}=\nu b_\pm(\nu),\quad
\forall\ \nu\in\RR\ .
\label{b-prop}
\eea

A second important operator is the  operator of topological charge
which can be written in terms of the fields as follows
\beq\label{Q}
Q=-\into dr\ \sk{e^\al\psi_-(r,\al)\psi_+(r,\al)
+e^{-\al}\psio_-(r,\al)\psio_+(r,\al)
}.
\eeq
In terms of the fermionic modes it has the form
\beq
Q
=-\intt d\nu\ {:}b_-(\nu)b_+(-\nu){:}=
\intt d\nu\ {:}b_+(\nu)b_-(-\nu){:}
\label{fer-ch}
\eeq
and is normalized in such a way that the charges
of the Rindler fermions $b_\pm(\nu)$ correspond to their indexes
\beq
{[}Q,b_\pm(\nu){]}=\pm b_\pm(\nu),\quad
\forall \nu\in\RR.
\label{ch-prop}
\eeq

In Rindler's parametrization the contours in the definition of
the conserved charges \r{charges} are
the straight rays $\a=\hbox{const}$. So in RRW we have
\beq\label{i1}
Q^J={1\over2}\int_0^\infty dr\left(\ee^{\a}J_y + \ee^{-\a}J_{\bar y}\right).
\eeq
Let us consider the charges given by this formula and compute them for the
conserved currents $J^0$ \r{iii2} and  $\tilde J^0$ \r{ii2}.
We obtain the result
\beq\label{xxxx}
I_n=(-1)^n\tilde I_n=
\intt d\theta\ e^{n\theta} {:}\Z_-(\theta)\Lambda_+(\theta){:}.
\eeq
Using formulas
\r{zz5} and \r{ZZ6}
we can observe that arbitrary non-vanishing matrix elements of the integrals
$I_n$ or $\tilde I_n$ in the Fermionic Fock space are divergent.
To avoid this divergence we consider
 the analytical continuation of
discrete index $n\to -i\la$ to the imaginary axis, where $\la\in\RR$.
In this case the charges given by  the currents  \r{ii1}, \r{ii2}
and \r{iii1}, \r{iii2} will produce  well
defined quantities. In contrast to \r{xxxx} the charges corresponding to
the currents  $J^0$ and  $\tilde J^0$ do not coincide.
We denote the ones corresponding to
neutral currents
\r{ii2} and
\r{iii2}
as $a_\la$, $\tilde a_\la$. In terms of the scattering data operators
or in Rindler fermions they have the form
\bea
a_\la&=&\intt d\theta
\ e^{i\la\theta}  {:} \Z_-(\theta)\Lambda_+(\theta){:}=
\intt d\nu\ \frac
{\Gamma\sk{\frac{1}{2}-i\nu}}
{\Gamma\sk{\frac{1}{2}+i(\la-\nu)}}
 {:}b_-(\la-\nu)b_+(\nu){:}\ ,
\label{ii6}\\
\tilde a_\la&=&\intt d\theta
\ e^{i\la\theta}  {:}\Lambda_-(\theta)\Z_+(\theta){:}=
\intt d\nu\ \frac
{\Gamma\sk{\frac{1}{2}-i(\la-\nu)}}
{\Gamma\sk{\frac{1}{2}+i\nu}}
 {:}b_-(\la-\nu)b_+(\nu){:}\ ,
\label{i6}
\eea
where normal ordering is defined with respect to fermionic vacuum vectors.

By comparing the formulas \r{i6} and \r{ii6} we conclude that the conserved
charges $a_\la$ and $\tilde a_\la$ are related to each other
by some complicated integral transform.
This integral transform can be described algebraically by extending
the algebra of the  operators $a_\la$ and $\tilde a_\la$.
This will be demonstrated in the Appendix \ref{app1}.

Using \r{stand} we see that
\beq\label{i15}
[a_\la,a_\mu]=[\tilde a_\la,\tilde a_\mu]=\la\delta(\la+\mu).
\eeq
and
this Heisenberg type commutation relation allows us to use these operators
for the bosonization.

\subsection{The bosonization and the screening currents}

Using \r{stand} we can obtain the commutation
relations:
\bea
\label{ii16}
[a_\la,\Z_-(\theta)]&=&e^{-i\la\theta} \Z_-(\theta),\qquad
[a_\la,\Lambda_+(\theta)]=-e^{-i\la\theta} \Lambda_+(\theta)\\
\label{i16}
[\tilde a_\la,\Lambda_-(\theta)]&=&e^{-i\la\theta} \Lambda_-(\theta),\qquad
[\tilde a_\la,\Z_+(\theta)]=-e^{-i\la\theta} \Z_+(\theta)
\eea
These formulas together with \r{i15} allow to bosonize the
operators $\Lambda_+(\theta)$ and $\Z_-(\theta)$ in terms of a free bosonic
field constructed from continuous bosons $a_\la$
and the operators
 $\Lambda_-(\theta)$ and $\Z_+(\theta)$
from the analogous free field constructed from the bosons $\tilde a_\la$.

This bosonization should
conserve all the properties of the Hilbert space $\CH_R^f$ and the action
of the operators $\Lambda_\pm(\theta)$, $\Z_\pm(\theta)$ on it.
It is clear that it is impossible to do this using only bosonic modes $a_\la$
because they carry the charge 0, while the Hilbert space
$\CH_R^f$ is naturally graded with respect to topological charge
operator:
\beq\label{HF-gr}
\CH_R^f=\bigoplus_{n\in\ZZ}\CH_{R,n}^f,\quad
\CH_{R,n}^f=\left\{\left. x\in \CH_{R}^f\ \right|\ Qx=nx\right\}
\eeq
Because of the formulas \r{zz5} and \r{ZZ6} the operators
$\Lambda_\pm(\theta)$ and $\Z_\pm(\theta)$ change the topological charge
\beq\label{tcch}
\Lambda_\pm(\theta),\ \Z_\pm(\theta)\ : \ \CH^f_{R,n}\to \CH^f_{R,n\pm1}\ .
\eeq
Note that in our normalization the topological charge operator
$Q$ coincides with the operators $-a_0\equiv -\tilde a_0$.

To conserve these properties of $\CH_{R}^f$ in the bosonization picture
we introduce a pair of zero mode operators $\Q$ and $\P$ which satisfy
the commutation relations
\beq\label{zero-mo}
[\P,\Q]=i
\eeq
and bosonic vacuum vectors $\bvac{n}$, $n\in\ZZ$ which are annihilated
by all nonnegative bosonic modes and are eigenstates of the operator $\P$
\beq\label{bFock}
a_\la\bvac{n}=0,\qquad
\hbox{$\la\geq 0$},\qquad \P\bvac{n}=n\bvac{n}\ .
\eeq
We identify $ \CH^f_{R,n}$ with bosonic space $ \CH^b_{R,n}$
generated from bosonic vacuum vector $\bvac{n}$
\beq\label{bFock1}
\int_{-\infty}^0
 f_n(\la_n) a_{\la_n} d\la_n\ldots \int_{-\infty}^0
 f_1(\la_1) a_{\la_1} d\la_1\
\bvac{n}\  ,
\eeq
 where the functions $f_i(\la)$ are analytical
functions in a neighborhood of
 ${\RR_-}$ except $\la=0$, where they can have a simple pole.

Then due to \r{i16} the operators $\Lambda_-(\theta)$ and $\Z_+(\theta)$
can be bosonized as follows:
\beq
\label{i17}
\Lambda_+(\theta)=\exp\sk{i\Q+
\vpint \frac{d\la}{\la}\ a_\la e^{i\la\theta}},\quad
\Z_-(\theta)=\exp\sk{-i\Q-
\vpint \frac{d\la}{\la}\ a_\la e^{i\la\theta}}.
\eeq
The integral under the exponent is understood as principal value integral
to exclude the singularity at zero:
\beq\label{vpint}
\vpint f(\la)\, d\la=\lim{\epsilon\to+0}\sk{\int_{-\infty}^{-\epsilon}
f(\la)\, d\la + \int_{\epsilon}^\infty f(\la)\, d\la}.
\eeq

We define the products of the operators
like \r{i17} to be $\zeta$-function regularized \cite{JM1,JMK}
\bea
&\exp\left(\intt d\la\ g_1(\la)\,a_\la\right)\ \cdot\
\exp\left(\intt d\mu\ g_2(\mu)\,a_\mu\right) = \nn\\
&\quad =\frac{e^{\gamma}}{2\pi}
\exp\left(\int_{\tilde C}{d\la\,\ln(-\la)\over2\pi i}\
c(\la)g_1(\la)g_2(-\la) \right)
\exp\left(\intt d\la\ (g_1(\la)+g_2(\la))\,a_\la\right)\ .
\label{normal}
\eea
where $c(\la)=\la$ and  $\gamma$ is Euler constant and
the contour $\tilde C$ is shown in  Fig.~1.
\bigskip

\unitlength 1.00mm
\linethickness{0.4pt}
\begin{picture}(121.00,20.00)
\put(17.00,15.00){\makebox(0,0)[cc]{0}}
\put(20.00,15.00){\makebox(0,0)[cc]{$\bullet$}}
\put(132.00,15.00){\makebox(0,0)[cc]{$+\infty$}}
\put(20.00,15.00){\line(1,0){100.33}}
\put(40.00,10.00){\line(1,0){80.33}}
\put(120.00,20.00){\line(-1,0){100.00}}
\put(30.00,5.00){\makebox(0,0)[cc]{Fig.~1.}}
\put(121.00,10.00){\vector(1,0){0.2}}
\put(100.00,10.00){\line(1,0){21.00}}
\put(20.00,10.00){\line(1,0){22.00}}
\put(20.67,10.00){\line(1,0){22.00}}
\put(20.17,15.00){\oval(15.00,10.00)[l]}
\end{picture}
\smallskip

Naturally there is an alternative way to bosonize the fermionic Fock space
$\CH_R^f$ using modes $\tilde a_\la$ and introducing the corresponding
zero mode operators $\tilde\Q$, $\tilde\P$, the
bosonic vacuum vectors and bosonic Fock spaces.
It is clear that these vacuum vectors are not a priori
the same as for bosons $a_\la$ because of the complicated commutation
relations between the bosons $a_\la$ and $\tilde a_\la$ (see Appendix
\ref{app1}). This alternative bosonization looks as follows
\beq
\label{ii17}
\Lambda_-(\theta)=\exp\sk{-i\tilde\Q-
\vpint \frac{d\la}{\la}\ \tilde a_\la e^{i\la\theta}},\quad
\Z_+(\theta)=\exp\sk{i\tilde \Q+
\vpint \frac{d\la}{\la}\ \tilde a_\la e^{i\la\theta}}\ .
\eeq

We also define the charged  currents in the momentum space
which correspond to the charged
conserved currents \r{ii1} and \r{iii1} by the requirement that
they relate the components of the  operators $\Z_\pm(\theta)$.
Let
\beq\label{I22}
E(\theta)=
{:}\Lambda_+(\theta+\pi i/2)
\Lambda_+(\theta-\pi i/2){:},\qquad
\tilde F(\theta)=
{:}\Lambda_-(\theta+\pi i/2)
\Lambda_-(\theta-\pi i/2){:}
\eeq
be the operator valued currents in the momentum space. Using vacuum
expectation values and Wick theorem we can prove the formulas
\bea
\Z_+(\theta)&=&-i\int_{C_1}du\ e^{u-\theta}\Z_-(\theta)E(u)+
i\int_{C_2}du\ e^{u-\theta}E(u)\Z_-(\theta)=
\Lambda_+(\theta+\pi i)+\Lambda_+(\theta-\pi i)\ ,\label{Z+Z-}\\
\Z_-(\theta)&=&-i\int_{C_1}du\ e^{u-\theta}\Z_+(\theta)\tilde F(u)+
i\int_{C_2}du\ e^{u-\theta}\tilde F(u)\Z_+(\theta)=
\Lambda_-(\theta+\pi i)+\Lambda_-(\theta-\pi i)\ ,\label{Z-Z+}
\eea
where the contour $C_1$ goes from $-\infty$ to $+\infty$
and  is above all the poles in the operator
product expansion $\Z_-(\theta)E(u)$ and $\Z_+(\theta)\tilde F(u)$
and the contour $C_2$ is also from $-\infty$ to $+\infty$
and  below all the poles in the OPE
$E(u)\Z_-(\theta)$ and $\tilde F(u)\Z_+(\theta)$.
Let us prove \r{Z+Z-}. It follows from the OPE
\bea
\Z_-(\theta)E(u)&=&{:}\Z_-(\theta)E(u){:}-
\frac{1}{2\pi i}\left[
\frac{\Lambda_+\sk{u-\frac{\pi i}{2}}}{\theta-u-\frac{\pi i}{2}}
-\frac{\Lambda_+\sk{u+\frac{\pi i}{2}}}{\theta-u+\frac{\pi i}{2}}
\right],\nn\\
E(u)\Z_-(\theta)&=&{:}\Z_-(\theta)E(u){:}-
\frac{1}{2\pi i}\left[
\frac{\Lambda_+\sk{u+\frac{\pi i}{2}}}{u-\theta-\frac{\pi i}{2}}
-\frac{\Lambda_+\sk{u-\frac{\pi i}{2}}}{u-\theta+\frac{\pi i}{2}}
\right].\nn
\eea
Now the formula \r{Z+Z-} follows from the trivial calculation of the
integrals. The second relation \r{Z-Z+} is proved analogously.

Using formulas \r{i17} and \r{ii17} we can write down
the bosonized expressions for the screening currents
\bea
E(u)&=&\exp\sk{2i\Q+2\vpint{d\la\over\la}\
e^{iu\la}\ch\sk{\pi\la/2}\ a_\la}\ ,\label{E-scr}\\
\tilde F(u)&=&\exp\sk{-2i\tilde\Q-2\vpint{d\la\over\la}\
e^{iu\la}\ch\sk{\pi\la/2}\ \tilde a_\la}\ ,\label{F-scr}
\eea
in terms of the bosons $a_\la$ for the current $E(u)$ and
$\tilde a_\la$ for $\tilde F(u)$.
Because the operators $a_\la$ and $\tilde a_\la$ do not form a closed
algebra the screening currents $E(u)$ and
 $\tilde F(u)$ also do not form the closed algebra. In the next subsection
we will define another pair of the screening currents
$\tilde E(u)$ and $F(u)$ such that the pairs $E(u)$, $F(u)$ or
$\tilde E(u)$, $\tilde F(u)$ do form the closed algebra both isomorphic to
the screening current algebra introduced in \cite{Lu1,KLP}.

\subsection{Another pair of screening currents and quantum Jost functions}

The commutation relations \r{YBt} have a smooth classical limit
when $\xi\to0$ and correspondingly $q'\to1$. In this limit these
commutation relations become  Poisson brackets for the elements
of monodromy matrices for classical SG model \cite{FTbook}.
The elements $Z'_{\ep}(\al)$ of the `monodromy' matrix associated with
the half-line
were introduced in \cite{Lu1}.
It was shown
 in \cite{Lu2}
that they are the
quantum analogs of classical Jost functions. These classical
objects can be written explicitly as path-ordered exponents
of SG connections in RRW and satisfy the Poisson bracket relation
\beq\label{PBrel}
\{\Z'_{\ep_1}(\al_1),\Z'_{\ep_2}(\al_2)\}=r_{\ep_1\ep_2}^{\ep'_1\ep'_2}
(\al_1-\al_2)     \Z'_{\ep'_2}(\al_2)\Z'_{\ep'_1}(\al_1)\ ,
\eeq
where $r_{\ep_1\ep_2}^{\ep'_1\ep'_2}
(\al)    $
is a classical trigonometric $r$-matrix
\beq\label{r-clas}
r_{\ep_1\ep_2}^{\ep'_1\ep'_2}
(\al) =\lim{\xi\to0}\frac{R_{\ep_1\ep_2}^{\ep'_1\ep'_2}
(\al,\xi)-1}{\pi i\xi}=\frac{1}{2}
\sk{\begin{array}{cccc}
   \th(\al/2) & &  &  \\
   &  -\cth(\al/2) & 2\sh^{-1}(\al) &  \\
   & 2\sh^{-1}(\al) &  -\cth(\al/2) &  \\
   &  &  & \th(\al/2)
\end{array}}
\eeq
obtained from the $R$-matrices \r{RR-mat}.
Note, that the scalar factor of the $R$-matrix \r{RR-mat} also contribute to
the classical $r$-matrix \r{r-clas}.

Unfortunately, a way to obtain
the quantum analog of the relations \r{PBrel}
\beq\label{PB-q}
\Z'_{\ep_1}(\al_1)\Z'_{\ep_2}(\al_2)=R_{\ep_1\ep_2}^{\ep'_1\ep'_2}
(\al_1-\al_2,\xi)     \Z'_{\ep'_2}(\al_2)\Z'_{\ep'_1}(\al_1)
\eeq
starting from SG Lagrangian is not known. Nevertheless, one can
formulate the properties of these operators which allows
to reconstruct them uniquely. These properties follow from the interpretation
of the operators $\Z_\pm(\theta)$ as the operators in $\CH_R$ which correspond
to the states in total Hilbert space $\CH$ of the model. Since
the operators $\Z'_\pm(\al)$ are related to the quantum Jost functions
and to the integrals of motion it is natural to require
their commutativity
with the operators $\Z_\ep(\theta)$
up to the phase
\beq\label{Z-Z'}
\Z_\ep(\theta)\Z'_\nu(\aa)=\ep\nu\  \phi(\theta-\aa)\
\Z'_\nu(\aa)\Z_\ep(\theta)
\eeq
where $\phi(\theta)$ is a yet unknown function
(see \r{ZFI-II} below). Because of the
relations \r{Z+Z-} and \r{Z-Z+} this requirement is equivalent to the
following
\beq\label{EF-Z'}
 E(u)\Z'_-(\aa)=-\Z'_-(\aa) E(u),\quad
\tilde F(u)\Z'_+(\aa)=-\Z'_+(\aa)\tilde F(u)
\eeq
These anticommutation relations  can be satisfied by the
following bosonization of the operators $Z'_\nu(\aa)$
\bea
\Z'_-(\al)=\exp\sk{i\Q/2
+\intt \frac{d\la}{\la}\ \frac{a_\la}{2\ch\,\pi\la/2}\ e^{i\la\al}},
\label{Z'+bos}\\
\Z'_+(\al)=\exp\sk{-i\tilde \Q/2
-\intt \frac{d\la}{\la}\ \frac{\tilde a_\la}{2\ch\,\pi\la/2}\
e^{i\la\al}}.
\label{Z'-bos}
\eea
Using this bosonization and also the rule of the normal ordering
\r{normal} we can observe  that the operators $\Z'_\pm(\al)$ are
related to the scattering data operators $\Lambda_\pm(\theta)$ as
follows
\beq\label{rel1}
\Lambda_{-\ep}(\aa)=g^{-1}
\Z'_\ep\sk{\al-\frac{i\pi}{2}}
\Z'_\ep\sk{\al+\frac{i\pi}{2}}={:}\Z'_\ep\sk{\al-\frac{i\pi}{2}}
\Z'_\ep\sk{\al+\frac{i\pi}{2}}{:}.
\eeq
The normalization constant is given in terms of double $\Gamma$-functions
(see Appendix \ref{app2} for the definition of these functions).
Note that formula \r{rel1} allows to identify the scattering
data operators $\Lambda_\pm(\theta)$ with
the generating functions of the local operators introduced in
the paper \cite{Lu1}. From
the formulas \r{Z'+bos} and \r{i17} we can easily find the
function $\phi(\theta)=\ctg\sk{\frac{\pi}{4}+\frac{\theta}{2i}}$.

The formulas in \r{rel1} are equalities in different bosonic
Fock spaces generated by the operators $a_\la$ and $\tilde a_\la$
respectively.  We can translate them in equal bosonic spaces
introducing another pair
of screening currents $F(u)$ and $\tilde E(u)$
\beq\label{scr2}
F(u)\equiv \Z_-(u),\quad \tilde E(u)\equiv \Z_+(u).
\eeq
These relations are given by the integral transforms
\bea
\Z_+'(\aa)&=&2e^{-2\gamma}\pi^{3/2}\int_{C'}du\  e^{\frac{u-z}{2}}
\left[e^{\frac{\pi i}{4}} \Z'_-(\aa)F(u)+
e^{-\frac{\pi i}{4}}F(u) \Z'_-(\aa) \right]\nn\\
&=&\frac{e^{-3\gamma/2}}{\sqrt2\pi}\int_{C'} du\
\Ga{\frac{1}{4}+\frac{u-\aa}{2\pi i}}
\Ga{\frac{1}{4}-\frac{u-\aa}{2\pi i}}
\ {:}\Z_-'(\aa)F(u){:},\label{rel+}
\eea
and
\bea
\Z_-'(z)&=&2e^{-2\gamma}\pi^{3/2}\int_{C'}du\  e^{\frac{u-z}{2}}
\left[e^{\frac{\pi i}{4}} \Z'_+(\aa)\tilde E(u)+
e^{-\frac{\pi i}{4}}\tilde E(u) \Z'_+(\aa) \right]\nn\\
&=&\frac{e^{-3\gamma/2}}{\sqrt2\pi}\int_{C'} du\
\Ga{\frac{1}{4}+\frac{u-\aa}{2\pi i}}
\Ga{\frac{1}{4}-\frac{u-\aa}{2\pi i}}
\ {:}\Z_+'(\aa)\tilde E(u){:},\label{rel-}
\eea
where in both formulas the contour $C'$
goes from $-\infty$ to $\infty$ along the real axis such that
\beq\label{contour'}
\Im\,\aa-\pi/2      < \Im\,u<  \Im\,\aa+\pi/2
\eeq
The proof of the
 fact that the relation \r{rel+} is equivalent to the relation
\r{rel1} for $\ep=-$ or vice versa \r{rel-} is equivalent to the
relation \r{rel1} for $\ep=+$ can be found in the Appendix \ref{app2}.

The fact that the second set of the screening currents for the quantum Jost
operators $\Z'_\pm(\aa)$ coincided with scattering data
 operators $\Z_\pm(\theta)$
is  specific to the FF point. But what is true in general
is that we have the closed algebra of screening currents either
for the pair $E(u)$, $F(u)$ or for the pair $\tilde E(u)$, $\tilde F(u)$.
Both of these algebras can be obtained from the commutation relations
\r{scr-com} using different Gauss decompositions of $L$-operators.

\setcounter{equation}{0}
\section{Algebra of screening currents}

As we already said in the Introduction one of the goals of this paper
is to explain the  algebraic structures which allows to describe
simultaneously two quantum group structures with different
parameter of deformations developing ideas of the paper \cite{Lu1}.
This algebra will be introduced and
explained in this section.

This is   a non-abelian
 algebra of screening currents which
can be defined using exact $S$-matrix of soliton-antisoliton
scattering in SG model \cite{ZZ,KLP}. We define this algebra for the value
of the of the renormalized coupling constant $1<\xi<\infty$ in so called
breatherless regime. We will demonstrate that the representation theory of
this algebra has smooth limit when $\xi\to1$, which corresponds to the FF
point of SG model. Using the bosonization we will show
that the intertwining operators
of the level 1 highest weight modules for the screening current algebra
coincide with the operators $\Z_\pm(\theta)$ and $\Z'_\pm(\theta)$ defined
in the previous section from analysis of massive Dirac fermions in RRW.

\subsection{$R$ and $S$ Matrices}

Consider the following $\R$-matrix.
\begin{eqnarray}
\R^+(u,\xi)&=&\tau^+(u)  \R (u,\xi),\quad
\R(u,\xi)\ =\ r (u,\xi)\overline \R (u,\xi)\ ,
 \label{R-mat}\\
\overline \R (u,\xi)&=&
\left(
\begin{array}{cccc}
1&0&0&0\\  0&b (u,\xi)&c (u,\xi)&0\\
0&\tilde c (u,\xi)&b (u,\xi)&0\\  0&0&0&1
\end{array}
\right)      \ ,     \nn\\
r (u,\xi)&=&{\Ga{\frac{1}{\xi}}\Ga{1+\frac{iu}{\pi\xi}}\over
        \Ga{\frac{1}{\xi}+\frac{iu}{\pi\xi}}}
\prod_{p=1}^\infty
{R _p(u,\eta) R _p(i\pi- u,\eta)
\over
R _p(0,\xi) R _p(i\pi,\xi)}\ , \label{inf-pr}\\
R _p(u,\eta)&=&{\Ga{\frac{2p}{\xi}+\frac{iu}{\pi\xi}}
               \Ga{1+\frac{2p}{\xi}+\frac{iu}{\pi\xi}}\over
        \Ga{\frac{2p+1}{\xi}+\frac{iu}{\pi\xi}}
        \Ga{1+\frac{2p-1}{\xi}+\frac{iu}{\pi\xi}}}\ ,\nn\\
b(u,\xi)&=&
{\sh\,\frac{u}{\xi}\over\sh\,\frac{u-\pi i}{\xi}}\ ,\quad
c(u,\xi)\ =\ -
{\sh\,\frac{i\pi}{\xi} \over\sh\,\frac{u-\pi i}{\xi}}\ ,\quad
\tau^+(u)=i\cth\sk{{u\over2}} . \nn
\end{eqnarray}
The scalar factor $r(u,\xi)$ has an integral representation
\beq\label{fac-r}
r(u,\xi)= \exp \left(2i\int_{0}^\infty{d\la\,\over\la}
\frac{\sh\,\la/2}{\sh\,\la}\ \frac{\sh\,(\xi-1)\la/2}{\sh\,\xi\la/2}
\ \sin\sk{\frac{\la u}{\pi}}\right),\quad -\pi<\Im\,u<\pi\ .
\eeq

The $R$-matrix \r{R-mat} differs from
the physical $S$-matrix which describe the soliton-antisoliton
scattering by the transformation
\beq\label{S-mat}
S(\b)=- r(\b,\xi)\overline S(\b),\quad
\overline S(\b)=
(\sigma_z\ot1)\overline \R(\b,\xi)(1\ot\sigma_z)
\eeq
which change the sign in front of the elements $b(\b,\xi)$.
In the classical limit
$\xi\to\infty$ such that $u/\xi$ is fixed the $R$-matrix $\R(u,\xi)$
goes to identity, while the `physical' $S$-matrix goes to
${\rm diag}(-1,1,1,-1)$.
On the other hand at the FF point $\xi\to1$
$S$-matrix becomes equal to $-1$ while
$R$-matrix \r{R-mat} to ${\rm diag}(1,-1,-1,1)$.

The matrix $\overline S(\b)$ can be written using a multiplicative
spectral parameter $z=e^{-\theta/\xi}$ and deformation parameter $q$
introduced by \r{s-mul}
\beq\label{S-mul}
\left(\begin{array}{cccc}
1&0&0&0\\
0&\frac{z-z^{-1}}{zq-z^{-1}q^{-1}}&\frac{(q-q^{-1})}{zq-z^{-1}q^{-1}}&0\\
0&\frac{(q-q^{-1})}{zq-z^{-1}q^{-1}}&\frac{z-z^{-1}}{zq-z^{-1}q^{-1}}&0\\
0&0&0&1
\end{array}\right).
\eeq

The physical $R$-matrix which describe the commutation
relations of quantum monodromy matrices in SG model \r{YBt}
and quantum Jost functions \r{PB-q}
can be similarly written in terms of the
matrix $\R(u,\xi)$
\beq\label{RR-mat}
\Rr(\aa)= r(\aa,\xi+1)\overline \Rr(\aa),\quad
\overline \Rr(\aa)=
(\sigma_z\ot1)\overline \R(-\aa,-\xi-1)(1\ot\sigma_z)
\eeq
Note that matrices $\overline S(\b)$ and $\overline \Rr(\aa)$
are related by the duality transformation
\beq\label{dual-tr}
\xi\to -\xi -1,\quad \theta\to-\al\ .
\eeq
The fact is that
the scalar factors of $S$ and $R$-matrices  are related by the same
transformation. To see this one should use simple identity
$$
\frac{\sh\,\la/2}{\sh\,\la}\sk{
\frac{\sh\,(\xi+1)\la/2}{\sh\,\xi\la/2}+
\frac{\sh\,(\xi-1)\la/2}{\sh\,\xi\la/2}}=1
$$
to rewrite the scalar factor $-r(\b,\xi)$ in the form
\beq\label{fac-r-n}
-r(u,\xi)= \exp \left(-2i\int_{0}^\infty{d\la\,\over\la}
\frac{\sh\,\la/2}{\sh\,\la}\ \frac{\sh\,(\xi+1)\la/2}{\sh\,\xi\la/2}
\ \sin\sk{\frac{\la u}{\pi}}\right).
\eeq
Now one can see that functions $r(u,\xi+1)$ and $-r(u,\xi)$ transform
to each other under \r{dual-tr}, although one should not think about
this transformation literally. The point is that the quantization of
the SG model is well defined
 for $0<\xi<\infty$, so in order to perform the dual transformation
\r{dual-tr} we should first go to $+\infty$ and then come back to the
negative axis from $-\infty$. During this path the properties of the
model itself change drastically.

\subsection{Algebra of screening currents}

Set $1<\xi<\infty$. Let
\beq
L(u,\xi)=\left(\begin{array}{cc}
L_{++}(u,\xi)&L_{+-}(u,\xi)\\ L_{-+}(u,\xi)&L_{--}(u,\xi)
\end{array}\right)
\label{L-op}
\eeq
be a quantum $L$-operator whose matrix elements are treated as generating
functions for the elements of the algebra given by the
commutation relations:
\begin{eqnarray}
R^+(u_1-u_2,\xi+c)L_1(u_1,\xi)L_2(u_2,\xi)&=&
L_2(u_2,\eta)L_1(u_1,\xi) R^+(u_1-u_2,\xi) \  , \label{RLL-univ}\\
\hbox{qdet} L(u)=L_{++}(u-i\tih)L_{--}(u)&-&L_{+-}(u-i\tih)L_{-+}(u)=1\ .
\label{qdet=1}
\eea
Let
\beq
L(u)
=\left(\begin{array}{cc} 1& f(u)\\0&1\end{array}\right)
\left(\begin{array}{cc}  k_1(u)&0\\ 0&k_2(u) \end{array}\right)
\left(\begin{array}{cc} 1&0\\ e(u)&1\end{array}\right)\ ,
\label{GL-univ}
\eeq
be the Gauss decomposition of the $L$-operator \r{L-op}.
This Gauss decomposition corresponds to the algebra of screening currents
$E(u)$ and $F(u)$ described in the previous section
at the value $\xi=1$. To obtain the algebra
related to the currents $\tilde E(u)$ and $\tilde F(u)$ we should start from
another Gauss decomposition
\beq
\tilde L(u)
=\left(\begin{array}{cc} 1& 0\\\tilde e(u)&1\end{array}\right)
\left(\begin{array}{cc}  \tilde k_1(u)&0\\ 0&\tilde k_2(u) \end{array}\right)
\left(\begin{array}{cc} 1&\tilde f(u)\\ 0&1\end{array}\right).
\label{GL-univ-al}
\eeq
The relation between the Gauss coordinates of both $L$-operators is
complicated enough and can be described on the level of the bosonization
of the $L$-operators by
the relations similar to those described in the Appendix \ref{app1}.
For the remainder
 of this paper we will work only with the operator \r{GL-univ}.

One can deduce from
 (\ref{RLL-univ}), (\ref{qdet=1}) that
$$
k_1(u)=(k_2(u+i\tih))^{-1}.
$$
Let
$$
h(u)=k_1\left(u\right)  k_2\left(u\right)^{-1}\ , \qquad
h'(u)=k_2\left(u\right)^{-1} k_1\left(u\right)=
\frac{\xi\sin\,(\pi/\xi)}{\xi'\sin\,(\pi/\xi')} h(u)
\ ,
$$
where by $\xi'$ we denote the combination $\xi+c$ of the parameter $\xi$
and the central element of the algebra $\AA$.

 The Gauss coordinates $e(u)$, $f(u)$ and $h(u)$ of
 the $L$-operator \r{L-op} satisfy the following commutation relations
 ($u=u_1-u_2$):
\beq
e(u_1)f(u_2)-f(u_2)e(u_1)
={\sh\,(i\pi/\xi')\over\sh\, (u/\xi')}h'(u_1)-
{\sh\,(i\pi/\xi)\over\sh\, (u/\xi)} h(u_2),\label{ef}
\eeq
\bea
\sh\sk{\frac{u+i\tih}{\xi}}h(u_1)e(u_2)-
\sh\sk{\frac{u-i\tih}{\xi}}e(u_2)h(u_1)
&=&\sh\sk{\frac{i\pi}{\xi}}
\{h(u_1),e(u_1)\},\label{he}\\
\sh\sk{\frac{u-i\tih}{\xi'}}h(u_1)f(u_2)-
\sh\sk{\frac{u+i\tih}{\xi'}}f(u_2)h(u_1)
&=&-\sh\sk{\frac{i\pi}{\xi'}}
\{h(u_1),f(u_1)\},\label{hf}\\
\sh\sk{\frac{u+i\tih}{\xi}}e(u_1)e(u_2)-
\sh\sk{\frac{u-i\tih}{\xi}}e(u_2)e(u_1)
&=&\sh\sk{\frac{i\pi}{\xi}}
\left(e(u_1)^2+e(u_2)^2\right),\label{ee-univ}\\
\sh\sk{\frac{u-i\tih}{\xi'}}f(u_1)f(u_2)-
\sh\sk{\frac{u+i\tih}{\xi'}}f(u_2)f(u_1)
&=&-\sh\sk{\frac{i\pi}{\xi'}}
\left(f(u_1)^2+f(u_2)^2\right),\label{ff-univ}
\eea
\beq
{\sh\sk{\frac{u-i\pi}{\xi'}}
\over
\sh\sk{\frac{u+i\pi}{\xi'}}}
h(u_1)h(u_2)
=
h(u_2)h(u_1)
{\sh\sk{\frac{u-i\pi}{\xi}}
\over
\sh\sk{\frac{u+i\pi}{\xi}}}
\ .\label{hh}
\eeq

In the next subsections we will describe the finite and infinite
dimensional representations of the algbera $\AA$. We will consider also the
tensor products of the representations where the action of the algebra $\AA$
is defined by
the following comultiplication structure compatible with the
commutation relations \r{RLL-univ}:
\bea
\Delta\,c&=&c'+c''=c\otimes1+1\otimes c\ ,\nn\\
\Delta^{\rm op} L(u,\xi)&=&L(u- i\pi c^{(2)}/4,\xi+c^{(2)})
\ {\dot\otimes}\ L(u+ i\pi c^{(1)}/4,\xi)
\label{comul-L-univ}\\
\Delta  \sk{L(u,\xi)}^{-1}&=&\sk{L(u+i\pi c^{(2)}/4,\xi)}
^{-1}
{\dot\otimes} \sk{L(u- i\pi c^{(1)}/4,\xi+c^{(1)})}^{-1},\nn
\eea
where the symbol $\dot\ot$ signifies the matrix tensor product
$$
\sk{A\dot\ot B}_{ij}=\sum_k A_{ik}\ot B_{kj}\ .
$$

The comultiplications of the  the Gauss coordinates
of $L$-operators $e(u,\xi)$, $f(u,\xi)$
and $h(u,\xi)$ are
\bea
\Delta e(u,\xi) &=&
e(u+ i\pi c^{(2)}/4,\xi)\ot 1+ \sum_{p=0} ^{\infty}(-1)^p
       \left(f(u+ i\pi c^{(2)}/4-i\tih,\xi)\right)^{p}\times\nn\\
&\times&
h(u+ i\pi c^{(2)}/4,\xi)\otimes
\left(e(u- i\pi c^{(1)}/4,\xi+c^{(1)}) \right)^{p+1}\ ,   \label{com-e-fu}\\
\Delta f(u,\xi)&=&
1\otimes f(u- i\pi c^{(1)}/4,\xi+c^{(1)}) +
\sum_{p=0} ^{\infty} (-1)^p
       \left(f(u+ i\pi c^{(2)}/4,\xi)\right)^{p+1}\ot\nn\\
&\otimes&
\tilde h(u- i\pi c^{(1)}/4,\xi+c^{(1)})
\left(e(u- i\pi c^{(1)}/4-i\tih,\xi+c^{(1)})
\right)^p\ ,   \label{com-f-fu}\\
\Delta h(u,\xi)&=&\sum_{p=0}^\infty (-1)^p
{\sin\,(\pi (p+1)/\xi)\over \sin\,(\pi/\xi)}
\left(f(u+ i\pi c^{(2)}/4-i\tih,\xi)\right)^{p}
h(u+ i\pi c^{(2)}/4,\xi) \ot\nn\\
&\ot&
h(u- i\pi c^{(1)}/4,\xi+c^{(1)})
\left(e(u- i\pi c^{(1)}/4-i\tih,\xi+c^{(1)}) \right)^p\ .
\label{comul-h}
\eea

\subsection{Finite-dimensional representations and the intertwining
operators}\label{3.3}

Let $e$, $f$ and $h$  generators of the algebra $U_{i\pi/\xi}(sl_2)$
with the commutation relations:
\beq\label{soot}
[h,e]=2e,\qquad [h,f]=-2f,\qquad [e,f]=
\frac{\sin\sk{\pi h/\xi}}{\sin\sk{\pi/\xi}}\ .
\eeq
The following formulas describe
the evaluation
homomorphism of the algebra $\Ael$ at $c=0$ onto the algebra
$U_{i\pi/\xi}(sl_2)$:
\bea
{\cal E}v_z \sk{e(u)}&=& -
{\sh\sk{i\pi/\xi}  \over
\sh\sk{\frac{u-z}{\xi}+\frac{i\pi(h-1)}{2\xi}}  }\, e= -
e\ {\sh\sk{i\pi\frac{\xi+1}{\xi}}  \over
\sh\sk{\frac{u-z}{\xi}+\frac{i\pi(h+1)}{2\xi}}  }\ ,
\nn\\
{\cal E}v_z \sk{f(u)}&=&-
{\sh\sk{i\pi/\xi} \over
\sh\sk{\frac{u-z}{\xi}+\frac{i\pi(h+1)}{2\xi}}  }\, f=
-f\ {\sh\sk{i\pi\frac{\xi+1}{\xi}}  \over
\sh\sk{\frac{u-z}{\xi}+\frac{i\pi(h-1)}{2\xi}}  }\ ,
\nn\\
{\cal E}v_z \sk{h(u)}&=&
\cos\sk{\frac{i\pi h}{\xi}} - \sh\sk{\frac{i\pi}{\xi}}\left[
\cth\sk{\frac{u-z}{\xi}+\frac{i\pi(h-1)}{2\xi}}\   ef\right.\nn\\
&-&\left.\cth\sk{\frac{u-z}{\xi}+\frac{i\pi(h+1)}{2\xi}}\ fe
\right].\label{eval-h-n}
\eea

Let
$V_n$ be $(n+1)$-dimensional $U_{q}(sl_2)$-module with a
 basis $v_k$,
$k=0,1,\ldots,n$ where  the operators $h$, $e$ and $f$
 act
according to the rules
\beq
h\,v_k=(n-2k)\,v_k,\quad e\,v_k=
\frac{\sin(\pi k \xi)}{\sin(\pi/\xi)}\,v_{k-1},\quad
f\,v_k=\frac{\sin(\pi (n-k)/\xi)}{\sin(\pi/\xi)}\,v_{k+1}\ .
\label{finite}
\eeq
Combining these formulas with the evaluation homomorphism we can construct
the level zero evaluation representations of the algebra $\AA$ in the
finite-dimensional space
$V_n$. In particular, in what follows we need the evaluation
representation of this algebra into two-dimensional space $V_1$. It is given
by the formulas
\bea
\pi_z\sk{e(u)}v_{+}&=&0,\quad \pi_z\sk{f(u)}
v_{-}\, =\, 0\ , \label{eval-0}\\
\pi_z\sk{e(u)}v_{-}&=& -
{\sh\sk{i \pi/\xi} \over \sh\sk{\frac{u-z}{\xi}}  }\
 v_{+}\ ,\quad
\pi_z\sk{f(u)}v_{+}\ =\ -
{\sh\sk{i \pi/\xi} \over \sh\sk{\frac{u-z}{\xi}}  }\
 v_{-}\ ,\label{eval-e-f1}\\
\pi_z\sk{h(u)}v_{\pm}&=&\cos\sk{\frac{\pi}
{\xi}}\mp\sh\sk{\frac{i\pi}{\xi}}
\cth\sk{\frac{u-z}{\xi}} v_{\pm}
= {\sh\sk{\frac{u-z\mp i\pi}{\xi}}
\over \sh\sk{\frac{u-z}{\xi}}  }
 v_{\pm}
\ .\label{eval-h1}
\eea
In these formulas we have identified $v_{+}=v_{0}$ and $v_{-}=v_{1}$.
Using formulas  \r{eval-0}--\r{eval-h1} we can define certain intertwining
operators between level one highest weight modules over the algebra $\AA$.

It was shown in \cite{KLP} that the algebra $\AA$ has the highest weight
representations at the value of the central element $c=1$ which can be
bosonized using one free continuous bosonic field.
We denote this representation space by the symbol $\H$ and will demonstrate
in the next subsection that at the FF point it coincides with the
bosonized version of the Hilbert
space of the free massive Dirac field in the RRW $\CH_R^b$.
In analogy with the group-theoretical  description of the quantum integrable
models on the infinite-dimensional lattice \cite{JM}
we define four types of the {\it twisted} intertwining operators
\begin{eqnarray}
Z'(z)&:& \H\to \H\otimes V_{1}\ ,\quad
Z^{\prime *}(z)\ :\ \H\ot V_{1} \to \H\ ,\nn
\\
Z^{*}(z)&:&V_{1}\otimes \H\to \H\ ,\quad
Z(z)\ :\  \H\to  V_{1}\ot \H\ .  \label{inteq0}
\end{eqnarray}
The algebra $\AA$ acts on the two-dimensional evaluation module $V_1$ by
the formulas \r{eval-0}--\r{eval-h1}.
We require that
these operators commute with the action of the algebra $\AA$ up to the
the involution \r{L-oper-rel}
\begin{eqnarray}
Z(z) \iota(x) &=& \Delta(x) Z(z)\ ,\quad
Z^{\prime *}(z)\Delta(x)\ =\  \iota(x)Z^{\prime *}(z)\ , \nn\\
Z^{*}(\b)\Delta(x)&=& \iota(x)Z^{*}(z)\ ,\quad
Z(z) \iota(x) \ =\  \Delta(x) Z(z)\ ,\quad
                          \label{inteq}
\end{eqnarray}
where $x\in\AA$.
Due to the dimension of the module $V_1$
the intertwining operators have two components which are defined as follows:
\bea
Z'(z) v  &=& Z'_+(z)v\otimes v_+ + Z'_-(z)v\otimes v_-\ ,\quad
Z^{\prime *}(z)(v\ot v_\pm)\ =\  Z^{\prime *}_\pm(z)v \ , \nn\\
Z^{*}(z)(v_\pm\otimes v)&=& Z^{*}_\pm(z)v \ ,\quad
Z(z) v  \ =\  v_+\ot Z_+(z)v + v_-\ot Z_-(z)v\ , \nn
\eea
where $v\in\H$.

Using the coalgebraic structure of the algebra $\AA$ we can rewrite the
defining relations \r{inteq} for the components of the intertwining
operators as commutativity with Gauss coordinates of $L$-operators.
For $x=h(u)$ in \r{inteq} we have
\bea
h(u)Z^*_-(\theta)h^{-1}(u)&=&
\frac{\sh\sk{\frac{u-\theta+5i\pi/4}{\xi}}}
     {\sh\sk{\frac{u-\theta+ i\pi/4}{\xi}}}Z^*_-(\theta)\ ,\label{51a}\\
h(u)Z_+(\theta)h^{-1}(u)&=&
\frac{\sh\sk{\frac{u-\theta+ i\pi/4}{\xi}}}
     {\sh\sk{\frac{u-\theta-3i\pi/4}{\xi}}}Z_+(\theta)\ ,\label{51b}\\
h(u)Z^{\prime*}_+(\al)h^{-1}(u)&=&
\frac{\sh\sk{\frac{u-\theta-5i\pi/4}{\xi+1}}}
 {\sh\sk{\frac{u-\theta- i\pi/4}{\xi+1}}}Z^{\prime*}_+(\al)\ ,\label{51c}\\
h(u)Z'_-(\al)h^{-1}(u)&=&
\frac{\sh\sk{\frac{u-\theta- i\pi/4}{\xi+1}}}
     {\sh\sk{\frac{u-\theta+3i\pi/4}{\xi+1}}}Z'_-(\al)\ .\label{51d}
\eea
For $e(u)$ we have
\bea
\sh\,\sk{\frac{i\pi}{\xi}} Z^*_+(\theta)&=&
\sh\,\sk{\frac{u-\theta+i\pi/4}{\xi}}
e(u)Z_-^*(\theta)+\nn\\
&+&
\sh\,\sk{\frac{u-\theta+5i\pi/4}{\xi}}
Z_-^*(\theta)e(u)\ ,\label{52a}\\
\sh\,\sk{\frac{i\pi}{\xi}} Z_-(\theta)&=&
\sh\,\sk{\frac{u-\theta-3i\pi/4}{\xi}}
e(u)Z_+(\theta)+\nn\\
&+&
\sh\,\sk{\frac{u-\theta+i\pi/4}{\xi}}
Z_+(\theta)e(u)\ ,\label{52b}\\
\{e(u),Z_+^{\prime*}(\theta)\}&=&\{e(u),Z_-'(\theta)\}\ =\ 0\ ,\label{52c}
\eea
and finally for $f(u)$
\bea
\sh\,\sk{\frac{i\pi}{\xi+1}} Z^{\prime*}_-(\al)&=&
\sh\,\sk{\frac{u-\al-i\pi/4}{\xi+1}}
f(u)Z_+^{\prime*}(\al)+\nn\\
&+&
\sh\,\sk{\frac{u-\al-5i\pi/4}{\xi+1}}
Z_+^{\prime*}(\al)f(u)\ ,\label{53a}\\
\sh\,\sk{\frac{i\pi}{\xi+1}} Z'_+(\al)&=&
\sh\,\sk{\frac{u-\al+3\pi/4}{\xi+1}}
f(u)Z'_-(\al)+\nn\\
&+&
\sh\,\sk{\frac{u-\al-i\pi/4}{\xi+1}}
Z'_-(\al)f(u)\ ,\label{53b}\\
\{f(u),Z_-^*(\theta)\}&=&\{f(u),Z_+(\theta)\}\ =\ 0\ .\label{53c}
\eea
Note that in all commutation relations for the operators $Z_\pm(\theta)$
or $Z^*_\pm(\theta)$ appear only the trigonometric functions with the periods
$2i\pi/\xi$ while in all those related to the operators $Z'_\pm(\al)$
or $Z^{\prime*}_\pm(\al)$ with the period $2i\pi/(\xi+1)$. This property is
encoded into the comultiplication rules \r{com-e-fu}--\r{comul-h}.

We did not write down all the relations following from \r{inteq}
 for the components of the
intertwining operators but only independent ones.  For example, the
relation \r{51a} is obtained by applying \r{inteq} to the vector
$v_-\ot v\in V_1\ot \H$. If we apply it to the vector
$v_+\ot v$ we obtain the relation
\bea
\sh\,\sk{\frac{2i\pi}{\xi}} Z^*_-(\theta)h(u)e(u-i\pi)&=&
\sh\,\sk{\frac{u-\theta+i\pi/4}{\xi}}
h(u)Z_+^*(\theta)-\nn\\
&-&
\sh\,\sk{\frac{u-\theta-3i\pi/4}{\xi}}
Z_+^*(\theta)h(u)\nn
\eea
which is
a consequence of \r{52a} and \r{he}. Nevertheless, the defining relations
\r{51a}--\r{53c} allow one to calculate some properties of the intertwining
operators.
For example, the commutation relations
\begin{eqnarray}
Z^*_{\nu_1}(\theta_1) Z^*_{\nu_2}(\theta_2)
&=&\rho(\theta_1-\theta_2)
\overline S_{\nu_1\nu_2}^{\nu'_1\nu'_2} (\theta_1-\theta_2,\xi)
Z^*_{\nu'_2}(\theta_2) Z^*_{\nu'_1}(\theta_1)\ , \nn \\
Z'_{\ep_2}(\al_2) Z'_{\ep_1}(\al_1)
&=&\rho'(\theta_1-\theta_2)
\overline R_{\ep_1\ep_2}^{\ep'_1\ep'_2} (\al_1-\al_2,\xi)
Z'_{\ep'_1}(\al_1) Z'_{\ep'_2}(\al_2)\ . \nn
\eea
can be proved using only \r{51a}-\r{53c} and supposing that
the operators $Z^*_-(\theta_1)$, $Z^*_-(\theta_2)$
commute up to some scalar factor and analogously  for
the operators $Z'_-(\al_1)$, $Z'_-(\al_2)$.

The defining relations \r{51a}--\r{53c} allow one to find the bosonization of
the intertwining operators from the bosonization of the screening current
algebra. Being specialized to the FF point $\xi=1$ the intertwining
operators will coincide with the operators $\Z_\pm(\theta)$,
and $\Z'_\pm(\al)$ constructed in the previous
section modulo shifts in the spectral parameters.
Also, all the scalar coefficients mentioned above can be fixed using
these bosonizations. We will do this in the next subsection.

\subsection{Bosonization of the screening operator algebra}

The description of  the infinite dimensional representations of the algebra
$\AA$ at non-zero value of the central element $C$ is divided into two steps.
The first step is to rewrite the commutation relations \r{ef}--\r{hh} in terms
of the total currents $E(u)$, $F(u)$ and $H(u)$:
\bea
e\sk{u-\fract{ic\tih}{4}}+e\sk{u-i\pi\xi-\fract{ic\tih}{4}}&=&
\xi\sin\,(\pi/\xi) E(u)\ ,\label{FDE}\\
f\sk{u+\fract{ic\tih}{4}}+f\sk{u-i\pi\xi'+\fract{ic\tih}{4}}&=&
\xi'\sin\,(\pi/\xi') F(u)\ ,\label{FDF}\\
h(u)&=&2\pi\xi'\ \sin(\pi/\xi')\ H\sk{u+\frac{i\pi\xi}{2}+\frac{i\pi c}{4}}
\ .\label{H-new}
\eea
We write the commutation relations for the total currents
in the form adequate for the category
of the highest weight representations:
\beq\label{EF}
[E(u),F(v)]=
\left[ \delta\sk{u-v-\frac{i\pi c}{2}}H\sk{v+\frac{i\pi(\xi+c)}{2}}
-
\delta\sk{u-v+\frac{i\pi c}{2}}H\sk{v-\frac{i\pi(\xi+c)}{2}}
\right],
\eeq
\bea
{ \Ga{\frac{1}{2}+\frac{1}{\xi}+\frac{i(u-v)}{\pi\xi}}
\over
\Ga{\frac{1}{2}-\frac{1}{\xi}+\frac{i(u-v)}{\pi\xi}} }
H(u)E(v)
&=&
E(v)H(u)
{ \Ga{\frac{1}{2}+\frac{1}{\xi}-\frac{i(u-v)}{\pi\xi}}
\over
\Ga{\frac{1}{2}-\frac{1}{\xi}-\frac{i(u-v)}{\pi\xi}} },
\label{HE}\\
{ \Ga{\frac{1}{2}-\frac{1}{\xi'}+\frac{i(u-v)}{\pi\xi'}}
\over
\Ga{\frac{1}{2}+\frac{1}{\xi'}+\frac{i(u-v)}{\pi\xi'}} }
H(u)F(v)
&=&
F(v)H(u)
{ \Ga{\frac{1}{2}-\frac{1}{\xi'}-\frac{i(u-v)}{\pi\xi'}}
\over
\Ga{\frac{1}{2}+\frac{1}{\xi'}-\frac{i(u-v)}{\pi\xi'}} },
\label{HF}\\
{ \Ga{1+\frac{1}{\xi}+\frac{i(u-v)}{\pi\xi}}
\over
\Ga{-\frac{1}{\xi}+\frac{i(u-v)}{\pi\xi}} }
E(u)E(v)
&=&
-E(v)E(u)
{ \Ga{1+\frac{1}{\xi}-\frac{i(u-v)}{\pi\xi}}
\over
\Ga{-\frac{1}{\xi}-\frac{i(u-v)}{\pi\xi}} },
\label{EE}\\
{ \Ga{1-\frac{1}{\xi'}+\frac{i(u-v)}{\pi\xi'}}
\over
\Ga{\frac{1}{\xi'}+\frac{i(u-v)}{\pi\xi'}} }
F(u)F(v)
&=&
-F(v)F(u)
{ \Ga{1-\frac{1}{\xi'}-\frac{i(u-v)}{\pi\xi'}}
\over
\Ga{\frac{1}{\xi'}-\frac{i(u-v)}{\pi\xi'}} },
\label{FF}
\eea
\bea
&{ \Ga{1+\frac{1}{\xi}+\frac{i(u-v)}{\pi\xi}}
\over
\Ga{1-\frac{1}{\xi}+\frac{i(u-v)}{\pi\xi}} }
{ \Ga{1-\frac{1}{\xi'}+\frac{i(u-v)}{\pi\xi'}}
\over
\Ga{1+\frac{1}{\xi'}+\frac{i(u-v)}{\pi\xi'}} }
H(u)H(v)
=\nn\\
&\qquad=H(v)H(u)
{ \Ga{1+\frac{1}{\xi}-\frac{i(u-v)}{\pi\xi}}
\over
\Ga{1-\frac{1}{\xi}-\frac{i(u-v)}{\pi\xi}} }
{ \Ga{1-\frac{1}{\xi'}-\frac{i(u-v)}{\pi\xi'}}
\over
\Ga{1+\frac{1}{\xi'}-\frac{i(u-v)}{\pi\xi'}} }.
\label{HH}
\eea

The commutation relations for the total currents \r{HE}--\r{HH} are written
in the form of equalities of the meromorphic functions without
poles and zeros \cite{DK,DKKP}. This means that the product
of the currents has the structure of poles and zeros defined
by the zeros and poles of the function which is in front of this product
in the commutation relations \r{HE}--\r{HH}. For example,
the product $E(u)E(v)$ has  poles at the points
$u=v-i\pi+i\pi\xi k$ and zeros at the points
$u=v+i\pi+i\pi\xi (k+1)$, $k\geq0$.

The Frenkel-Ding \cite{DF} formulas \r{FDE} and \r{FDF}
 for the total currents can be inverted
solving the Riemann-Hilbert problem associated with the strips
of the widths $\pi\xi$ and $\pi\xi'=\pi(\xi+c)$ for the currents
$E(u)$ and $F(u)$ respectively \cite{KLP}
\bea
e(u)&=&\sin\,\pi/\xi\ \int_{C} {dv\over2\pi i}\
 { E(v)\over\sh\,\frac{u-v+ ic\tih/4}{\xi}}\ ,\label{2e}\\
f(u)&=&\sin\,\pi/\xi'\  \int_{C'}
{dv\over2\pi i}\  { F(v)\over\sh\,
\frac{u-v- ic\tih/4}{\xi'}}\ ,\label{2f}
\eea
where the contour $C'$ goes from $-\infty$ to $+\infty$,
the points $u+ic\tih/4+ik\pi\xi'$ ($k\geq0$) are above the contour and
the points $u-ic\tih/4-ik\pi\xi'$ ($k\geq0$) are below the contour.
The contour $C$ also goes from $-\infty$ to $+\infty$ but
the points $u-ic\tih/4+ik\pi\xi$ ($k\geq0$) are above the contour and
the points $u+ic\tih/4-ik\pi\xi$ ($k\geq0$) are below the contour.

The second step is the bosonization of the currents
$E(u)$, $F(u)$ and $H(u)$.
To describe the symmetries of the SG model we need the bosonization
of the algebra $\AA$ at the value of the central element $c=1$.
To construct this bosonization we define the continuous Heisenberg
operators  $a_\la$ which satisfy the commutation relations
\beq\label{gen-case}
[a_\la,a_\mu]=\la\,
\frac{\sh\,\frac{\pi\la}{2}}{\sh\,\pi\la}\,
\frac{\sh\,\frac{\pi\la(\xi+1)}{2}}{\sh\,\frac{\pi\la\xi}{2}}
\delta(\la+\mu)=c(\la)\delta(\la+\mu)\ .
\eeq
The commutation relations of the currents
$E(u)$, $F(u)$ and $H(u)$ are satisfied by the operators
\bea
E(u)&=&\exp\sk{2i\Q+2\vpint{d\la\over\la}\ e^{iu\la}\ch\sk{\pi\la/2}\
a_\la}
\ ,\label{E-bos}\\
F(u)&=&\exp\sk{-\frac{2i\xi}{\xi+1}\ \Q-2
\vpint{d\la\over\la}\ e^{iu\la}
\frac{
\ch\sk{\pi\la/2} \sh\sk{\pi\la\xi/2}}
{\sh\sk{\pi\la(\xi+1)/2}}
\ a_\la}
\ ,\label{F-bos}\\
H(u)&=&\exp\sk{\frac{2i}{\xi+1}\ \Q+
\vpint{d\la\over\la}\ e^{iu\la}
\frac{ \sh\sk{\pi\la}}
{\sh\sk{\pi\la(\xi+1)/2}}
\ a_\la}
\ ,\label{H-bos}
\eea
To verify this statement we should use the normal ordering rule given by
\r{normal} with the function $c(\la)$ specified in \r{gen-case} and
formulas given in the Appendix \ref{app2}.

Note that at the FF point the Heisenberg operators
$a_\la$ become the same as \r{i15} of the nonlocal
integrals of motion \r{ii6} so the  bosonization of the current $E(u)$
coincides with the bosonization \r{E-scr}, the bosonization
of the current $F(u)$ coincides with the bosonization of the
scattering operator $\Z_-(u)$ and $H(u)$ with
$\Lambda_+(u)$ (cf. \r{i17}). The commutation relations \r{HE}--\r{HH}
become in this case
$$
[H(u),E(v)]=[E(u),E(v)]=\{H(u),F(v)\}=\{F(u),F(v)\}=\{H(u),H(v)\}=0\ .
$$
The commutation relation
\r{EF} being multiplied by $e^{u-v}$ and integrated over the parameter
$u$ becomes the relation \r{Z+Z-}
which relates the components of the scattering
data operators $\Z_\pm(\theta)$.

The formulas \r{51a}--\r{53c} allow to bosonize the components of the
intertwining operators. It is given by the following formulas
\cite{Lu1,KLP}
\bea
Z_+(\theta)&=&
\exp\left(-i\Q-\vpint {d\la\over\la}\ \ee^{-i\la (\theta+\pi i/2)}\
a_\la\right)
\ ,\label{Psi-}\\
Z_-(\theta) &=& \int_{C}
{du\over2\pi}\  e^{(u-\theta)/\xi}   \left[(q)^{1/2} E(u)
Z_+(\theta)-(q)^{-1/2}Z_+(\theta)E(u)\right]\ ,\label{Psi+}\\
Z^*_\pm(\theta)&=&Z_{\mp}(\theta-i\tih)\ ,\label{Psidual}\\
Z'_-(\al)&=&
\exp\sk{\frac{i\xi}{\xi+1}\ \Q+
\vpint{d\la\over\la}\ e^{-i\la(\al-\pi i/2)}
\frac{ \sh\sk{\pi\la\xi/2}}
{\sh\sk{\pi\la(\xi+1)/2}}
\ a_\la}
 ,\label{Phi-}\\
Z'_+(\al) &=& \int_{C'}
{du\over2\pi}\   e^{(u-\al)/(\xi+1)} \left[(q')^{1/2}
Z'_+(\al)F(u)-(q')^{-1/2}F(u)Z'_+(\al)\right]\ , \label{Phi+}\\
Z^{'*}_\pm(\al)&=&Z'_{\mp}(\al+i\tih)\ , \label{Phidual}
\eea
where $q$ and $q'$ are given by \r{s-mul} and \r{q'-def} respectively
and the contour $C'$ goes from $-\infty$ to $+\infty$ along the real
axis leaving
the points $z+i\tih/2+ik\pi(\xi+1)$ ($k\geq0$) above the contour and
the points $z-i\tih/2-ik\pi(\xi+1)$ ($k\geq0$) below the contour.
The contour $C$ also goes from $-\infty$ to $+\infty$ but
the points $z-i\tih/2+ik\pi\xi$ ($k\geq0$) are above the contour and
the points $z+i\tih/2-ik\pi\xi$ ($k\geq0$) are below the contour.

The formulas \r{Psi-}, \r{Psi+} and \r{Phi-}, \r{Phi+} demonstrate that
at the FF point
the operators $Z_\pm(\theta)$ coincide with the operators
$\Z_\mp(\theta+\pi
i/2)$ modulo certain normalization constants. The same is true for the
relation between operators $Z'_\pm(\al)$ and $\Z'_\pm(\al-\pi i/2)$.

The second remark concerns the form of the contour $C$ in the relation
\r{Psi+}. The form of this contour is shown on the Fig.~2.
\bigskip
\bigskip

\unitlength 1mm
\linethickness{0.4pt}
\begin{picture}(130.67,38.67)
\bezier{148}(10.33,18.67)(42.67,18.67)(47.33,17.33)
\bezier{40}(47.33,17.33)(52.33,15.67)(56.33,12.33)
\bezier{44}(56.33,12.33)(59.67,9.33)(66.00,7.67)
\bezier{68}(66.00,7.67)(77.00,5.33)(83.00,7.00)
\bezier{84}(83.00,7.00)(92.67,12.33)(84.00,16.33)
\bezier{92}(84.00,16.33)(73.67,20.33)(62.00,24.00)
\bezier{80}(62.00,24.00)(53.33,28.67)(62.00,33.00)
\bezier{92}(62.00,33.00)(73.67,37.33)(83.00,33.33)
\bezier{48}(83.00,33.33)(91.00,30.00)(91.33,26.67)
\bezier{32}(91.33,26.67)(93.00,22.33)(96.33,22.00)
\bezier{76}(96.33,22.00)(105.67,21.00)(115.33,21.00)
\put(130.33,21.00){\vector(1,0){0.2}}
\put(115.33,21.00){\line(1,0){15.00}}
\put(79.33,30.00){\makebox(0,0)[cc]{$\bullet$}}
\put(79.33,11.33){\makebox(0,0)[cc]{$\circ$}}
\put(75.50,13.00){\makebox(0,0)[rc]{$\theta_1-i\tih/2$}}
\put(75.50,28.00){\makebox(0,0)[rc]{$\theta_1+i\tih/2$}}
\put(130.67,25.67){\makebox(0,0)[cc]{$+\infty$}}
\put(10.33,26.00){\makebox(0,0)[cc]{$-\infty$}}
\put(10.00,9.33){\makebox(0,0)[cc]{Fig.~2.}}
\put(79.33,38.67){\makebox(0,0)[cc]{$\circ$}}
\put(75.50,38.67){\makebox(0,0)[rc]{$\theta_1-i\pi/2+i\pi\xi$}}
\put(79.33,0.67){\makebox(0,0)[cc]{$\bullet$}}
\put(75.50,0.67){\makebox(0,0)[rc]{$\theta_1+i\pi/2-i\pi\xi$}}
\put(85.33,30.00){\makebox(0,0)[cc]{$\bullet$}}
\put(92.33,30.00){\makebox(0,0)[lc]{$\theta_2+i\pi/2$}}
\put(85.33,0.67){\makebox(0,0)[cc]{$\bullet$}}
\put(91.33,0.67){\makebox(0,0)[lc]{$\theta_2+i\pi/2-i\pi\xi$}}
\put(85.33,-2.67){\makebox(0,0)[rc]{$\vdots$}}
\put(79.33,-2.67){\makebox(0,0)[rc]{$\vdots$}}
\put(79.33,43.67){\makebox(0,0)[rc]{$\vdots$}}
\end{picture}
\bigskip
\bigskip

\noindent We can see from this picture that
in the limit to the FF point $(\xi\to1)$ there is a double pinching of
the integral which leads to the relation \r{Z+Z-} where  the integrals are
calculated as the residues in the points $u=\theta\pm\pi i/2$.
Moreover, this figure demonstrates
that the product $Z_+(\theta_1)Z_-(\theta_2)$
has the pole when $\theta_1\to\theta_2+\pi i$ because of the pinching of  the
contour in the integral representation of this quantity.
The origin of this pole due to pinching does not yield the restriction on
the domain of the definition  of this product and
$Z_+(\theta_1)Z_-(\theta_2)$ is an meromorphic function of the variable
$\theta_1-\theta_2$ in the domain $\Im(\theta_1-\theta_2)>-\pi i/2$ with a
simple pole in the point $\theta_1\to\theta_2+\pi i$.

Using standard techniques \cite{Lu1,KLP} we can find the properties
of the intertwining operators:
\begin{eqnarray}
Z_{\nu_1}(\theta_1) Z_{\nu_2}(\theta_2)
&=&S_{\nu_1\nu_2}^{\nu'_1\nu'_2} (\theta_1-\theta_2,\xi)
Z_{\nu'_2}(\theta_2) Z_{\nu'_1}(\theta_1)\ , \label{ZFII} \\
Z'_{\ep_2}(\al_2) Z'_{\ep_1}(\al_1)
&=& R_{\ep_1\ep_2}^{\ep'_1\ep'_2} (\al_1-\al_2,\xi)
Z'_{\ep'_1}(\al_1) Z'_{\ep'_2}(\al_2)\ , \label{ZFI} \\
Z_{\nu}(\theta) Z'_{\ep}(\al) &=&\nu\ep\,\tg\left(
{i(\theta-\al)\over2} -{\pi\over4}
 \right)   Z'_{\ep}(\al)
Z_{\nu}(\theta)\ , \label{ZFI-II}\\
\label{complet}
\sum_{\ep=\pm} Z^{\prime*}_{\ep}(\al)Z'_{\ep}(\al)
&=&g'(\xi)\ \id\ ,\\
\label{orthI}
Z'_{\ep_1}(\al)Z^{\prime*}_{\ep_2}(\al)
&=&g'(\xi)\delta_{\ep_1\ep_2}\ \id\ ,\\
\label{orthII}
Z_{\ep_1}(\theta_1)Z^{*}_{\ep_2}(\theta_2)
&=&{g(\xi)\delta_{\ep_1\ep_2}\, \id
\over \theta_1-\theta_2} + o(z_1-z_2)\ ,
\end{eqnarray}
where  the $S$ and $R$-matrices is given by \r{S-mat} and \r{RR-mat}
respectively
and the   normalization constants $g(\xi)$, $g'(\xi)$ can be expressed
through double $\Gamma$-functions using the formulas given in the Appendix
\ref{app2} \cite{KLP}.

\setcounter{equation}{0}
\section{Angular quantization}

Before starting this section we would like to fix the terminology and
explain what we mean by the angular quantization in the context of
integrable quantum  field theory.
By this term we mean the possibility to represent the states and
operators in the total Hilbert space of the model associated with total
space-time as some operators acting in the Hilbert space associated
with RRW.
So, considering the free fermion in RRW in the second section
we did not really consider the angular
quantization but did only some preliminary work. The angular quantization
of SG model  will be considered in this section. But first
we would like to recall the angular quantization in lattice
integrable models inspired by Baxter's corner transfer matrix method.

\subsection{Angular quantization on the lattice}

In a series of papers, see e.g. \cite{XXZ,JM} the precise mathematical
description of anti-ferroelectric XXZ model in thermodynamic limit was
developed in terms of representation theory of quantum affine
Lie algebra $U_q(\widehat{sl}_2)$ with the real deformation parameter
$-1<q<0$. This description, based on Baxter's
corner transfer matrix method, looks as follows.

The total Hilbert space of the theory is identified with the space
of endomorphisms ${\rm End}(\Lambda_0\oplus\Lambda_1)$ of direct
sum of the level one irreducible $U_q(\widehat{sl}_2)$ modules with (complex
linear) scalar product given by the natural prescription
\beq\label{Ksc}
(A,B)=\Tr_{\Lambda_0\oplus\Lambda_1}AB\ .
\eeq
Two components of degenerated vacuum are identified, up to the constant,
with $(-q)^{D^{(1)}}$, where $D^{(1)}$
is principal gradation operator for quantum
affine algebra, multiplied by the projection to $\Lambda_i$.

The representation theory of $U_q(\widehat{sl}_2)$ provides two
types of operators
\beq\label{vertex}
\Phi(\zeta): \Lambda_i\to \Lambda_{1-i}\otimes V_\zeta,\quad
\Psi^*(\zeta): V_\zeta\otimes\Lambda_i\to \Lambda_{1-i},\quad
i=0,1
\eeq
which commute with the action of
 $U_q(\widehat{sl}_2)$.
Here $V_\zeta$ is a two-dimensional representation of  $U_q(\widehat{sl}_2)$
with basis $v_\pm$ evaluated at the point $\zeta^2$.

The transfer matrix $T(\zeta)$ of the theory acts on the state
$A\in{\rm End}(\Lambda_0+\Lambda_1)$ as
\beq\label{trans}
T(\zeta)\cdot A=\sum_{\ep=\pm}\Phi_\ep(\zeta)\ A\ \Phi_{-\ep}(\zeta)
\eeq
and the eigenvectors of the transfer matrix are described in terms of the
second type intertwining operators:
\beq\label{st-lat}
|\xi_n,\ldots,\xi_1\rangle_{\ep_n,\ldots,\ep_1;(i)}=
c(n)\Psi^*_{\ep_n}(\xi_n)\ldots \Psi^*_{\ep_1}(\xi_1)
(-q)^{D^{(i)}}\ .
\eeq

The local spin operators $\sigma_n^\pm$ acting on the $n$th
cite of the lattice can be described in terms of operators $\Phi(\zeta)$.
Due to the definition of  scalar product it gives the expressions
of the correlation functions of finite products of operators  $\sigma_n^\pm$
and of the form-factors of a local operator in terms of traces
of products operators $\Phi(\zeta)$ and $\Psi^*(\xi)$ in the Fock
space $\Lambda_0\oplus\Lambda_1$ \cite{JM}. Moreover, the adjoint
(in a sense of Hopf algebra) action of $U_q(\widehat{sl}_2)$ equips
the space of states with a structure of level 0
$U_q(\widehat{sl}_2)$-module, such that $n$-particle states form
$n$-fold tensor products of the two-dimensional representations
of $U_q(\widehat{sl}_2)$.

\subsection{Angular quantization in the 2d field theory}

A counterpart of the CTM  ideology in the integrable models
of the 2d quantum
field theory
in the infinite volume
looks as follows \cite{BrL}.

Let $\CH_R$ be a Hilbert space of canonical quantization of a theory in
the RRW, where boost operator $K=-i\partial_\al$
 is considered as Hamiltonian. Here $\partial_\al$ is
the differentiation with respect to the angular `time' or,
what is the same, with respect of the spectral parameters (see \r{stpr}).

The total Hilbert space $\CH$ of the model is supposed to be
a properly defined subspace of ${\rm End}\,\CH_R$
with the scalar product $(A,B)=\Tr_{\CH_R}A\cdot B$. The vacuum state in
$\CH$ is
identified with the operator $e^{\pi K}$  in $\CH_R$
and the definition of the transfer matrix refers
to certain quantum version of Jost functions \cite{Lu1}, $Z'_\pm(\al)$
(here $\al$ is the spectral parameter)
whose precise construction on the quantum level is not known.
On the classical level these objects in SG theory were introduced
by S.Lukyanov in \cite{Lu2} using zero curvature
representation of SG equation in RRW: $[\partial_r-A_r,
\partial_\al-A_\al]=0$.

The asymptotic states $|\theta_1,\ldots,\theta_n\rangle_{\ep_1,\ldots,\ep_n}$
are presented by the
products of the operators
\beq\label{states}
|\theta_1,\ldots,\theta_n\rangle_{\ep_1,\ldots,\ep_n}=
Z^*_{\ep_1}(\theta_1)\ldots Z^*_{\ep_n}(\theta_n)e^{\pi  K},
\eeq
analytically
continued to the real line,
where $Z^*_{\pm}(\theta)$ are certain
operators acting in the RRW Hilbert
space $\CH_R$. They can be represented by the bosonized
expressions \r{Psi-}-\r{Psidual}.
The conjugated states are given by the product of the
operators $Z_\pm(\theta)=Z^*_\mp(\theta+\pi i)$:
\beq\label{states-co}
_{\ep_1,\ldots,\ep_n}\langle\theta_1,\ldots,\theta_n|=e^{\pi K}
Z_{\ep_1}(\theta_1)\ldots Z_{\ep_n}(\theta_n).
\eeq

Analogously to the lattice case, one can assume that any local operator
in the theory can be presented in this language in terms of left and
right multiplications of certain combinations of the operators
$Z'_{\ep}(\al)$ and thus form-factor of operator $O$ can be given by some
trace formula
\beq\label{tracecon}
\lvacph O |\theta_1,\ldots,\theta_n\rangle_{\ep_1,\ldots,\ep_n}=
\Tr_{\CH_R}\sk{e^{2\pi K}\tilde O
Z^*_{\ep_1}(\theta_1)\ldots Z^*_{\ep_n}(\theta_n)
},
\eeq
where $\tilde O$ is some operator acting in $\CH_R$ and corresponding to the
original operator $O$. The problem to find an expression for the operator
$\tilde O$ in terms of the quantum Jost operators $Z'_\pm(\al)$ is a
complicated problem and has no general solution for the arbitrary operator
$O$ although for some simple operators it can be solved by comparing the form
factors obtained in the framework of the bootstrap program with those obtained
by means of the formula \r{tracecon} (see \cite{Lu1,NPT} for the simplest
examples in case of the $SU(2)$-invariant Thirring model).
We understand the trace in \r{tracecon} as properly regularized
to produce the known form factor formulas in SG theory
given in \cite{Sbook} (see the paper
\cite{Ni} for the alternative formulation of a continuum analogue of the
 Baxter corner matrix method).

The possibility to present the matrix element
$\lvacph O |\theta_1,\ldots,\theta_n\rangle_{\ep_1,\ldots,\ep_n}$
as a trace \r{tracecon}, the relation \r{Psidual} and the fact that the
operators $Z_\pm(\theta)$ commute with the operators $\tilde O$ up to
numbers related to the locality index \cite{Lu1} allows to demonstrate
easily the crossing symmetry of these matrix elements. We have
\bea\label{crossing}
&_{\ep'_1,\ldots,\ep'_{n'}}\langle\theta'_1,\ldots,\theta'_{n'}|
 O |\theta_1,\ldots,\theta_n\rangle_{\ep_1,\ldots,\ep_n}=\nn\\
&\qquad=
\Tr_{\CH_R}\sk{e^{2\pi K}
Z_{\ep'_1}(\theta'_1)\ldots Z_{\ep'_{n'}}(\theta'_{n'})
\tilde O
Z^*_{\ep_1}(\theta_1)\ldots Z^*_{\ep_n}(\theta_n)
}=\nn\\
&\qquad=\ \
\lvacph O |\theta_1,\ldots,\theta_{n},
\theta'_1-i\pi,\ldots,\theta'_{n'}-i\pi\rangle
_{\ep_1,\ldots,\ep_{n},-\ep'_1,\ldots,-\ep'_{n'}}\ .
\eea

Using the trace formulas we can also verify the completeness of the
space of states \r{states} and \r{states-co} with respect to the scalar
product given by the trace over RRW Hilbert space $\CH_R$.
First of all we observe that the matrix element \r{tracecon} of the unity
operator vanishes identically because after substitution of the integral
representations of the operators $Z_+^*(\theta)$ \r{Psi+} in \r{tracecon}
we obtain the integral with the integrand being the total difference which
leads to the vanishing of the integral \cite{NPT}. On the other hand the
pairing of the states \r{states} and \r{states-co} does not vanish
identically but is proportional to some combinations of the $\delta$-functions.
In particular, the simplest pairing of the one-particle states is equal to
$$
_{\ep'}\langle\theta'|\theta\rangle_{\ep}=\delta_{\ep\ep'}\delta(\theta-\theta').
$$
The delta-function in this formula appears because the trace
$\Tr_{\CH_R}\sk{e^{\sigma K}
Z_{\ep'}(\theta')Z^*_{\ep}(\theta)
}$ has two simple poles in the points $\theta'=\theta$ due to \r{orthII} and
in the point $\theta=\theta'+\sigma + 2\pi i$ due to the trace properties.
When the parameter
$\sigma$ tends to the value $-2\pi i$ these two poles form the
$\delta$-function (see \cite{JM} for the detailed description of this
mechanism in a case of lattice integrable models). We would like to note
here that the
same mechanism is responsible for the fact that form factors of the
local operators satisfy the annihilation axiom \cite{Sbook}. For XXZ
model this fact was established in \cite{P}.

These are the general features of the angular quantization approach in the 2d
integrable field theory. In order for the angular quantization approach
be the self-consistent, in particular the traces \r{tracecon}
satisfy all the axioms in the form-factor approach \cite{Sbook},
 the operators $Z^*_\pm(\theta)$ and $Z'_\pm(\al)$ should satisfy the
properties \r{ZFII}--\r{orthII}. Since the representation theory of the
algebra $\AA$ contains the operators which satisfy such properties we claim
that this algebra is the dynamical symmetry algebra of the SG model in the
sense claimed in \cite{XXZ} for the quantum XXZ model.

Using the properties of the operators $Z'_\pm(\al)$ we can find
representations of the commutation relations \r{YBt} for the quantum
monodromy matrices and interpret its trace as the generating function of
the local integrals of motion through the asymptotical expansion.
This will be done in the next subsection. Moreover, we can define
appropriate adjoint action of the algebra $\AA$ onto the Hilbert space of
the SG model $\CH$ which describes the known symmetries of this space of
states related to the  quantum affine algebra $U_q(\widehat{sl}_2)$
\cite{BL} and
interpret these symmetries
as level zero action of the algebra $\AA$ in the Hilbert space of states.
In the last subsection
we will demonstrate that these symmetries
being specialized to the FF point become the symmetries governed by the
classical affine algebra at level zero and associated with the strip
\cite{KLPST}.

\subsection{Properties of the monodromy matrix in SG model}

In this and next subsections we will understand by the operators
$Z^*_\pm(\theta)$, $Z_\pm(\theta)$, $Z'_\pm(\al)$ and $Z^{'*}_\pm(\al)$
the intertwining operators of the algebra $\AA$ which satisfy the properties
\r{ZFII}--\r{orthII}.

A monodromy matrix of the model acting on any
state $X_k\in \CH_k$ of the total Hilbert space $\CH=
\CH_0\oplus\CH_1$, where $k=0,1$ corresponds
to the subspaces of $\CH$ of the even and odd number
of particles respectively,  is defined as follows
\beq\label{monod1}
{\cal T}_{\ep\ep'}(\aa)\cdot X_k=
\sk{g'(\xi)}^{-1}\ \ep^k Z'_\ep(\aa)\cdot X_k\cdot Z'_{-\ep'}(\aa),\quad
k=0,1\ .
\eeq
The commutation relations \r{ZFI} allow to find the commutation relation
for this matrices:
\beq\label{com-mon}
\Rr(\aa_1-\aa_2,\xi){\cal T}_1(\aa_1){\cal T}_2(\aa_2)=
{\cal T}_2(\aa_2){\cal T}_1(\aa_1)
\Rr(\aa_1-\aa_2,\xi)\
\eeq
which coincides with \r{YBt}.

The trace of the monodromy matrix or the transfer matrix $T(\aa)$ is
\beq\label{tr-monod}
T(\aa)\cdot X_k=\sk{g'(\xi)}^{-1}\
\sum_{\ep=\pm}\ep^k Z'_\ep(\aa)\cdot X_k\cdot Z'_{-\ep}(\aa)\ .
\eeq
The inverse transfer matrix is given in terms of the operators
$Z^{\prime*}_\pm(\aa)$:
\beq\label{tr-monod-in}
T^{-1}(\aa)\cdot X_k=\sk{g'(\xi)}^{-1}\
\sum_{\ep=\pm}\ep^k Z^{\prime*}_\ep(\aa)
\cdot X_k\cdot Z^{\prime*}_{-\ep}(\aa)\ .
\eeq
The fact that operators \r{tr-monod} and \r{tr-monod-in} are inverse to
each other is a direct consequence of the properties \r{complet} and
\r{orthI}. The same properties allow to prove that the physical vacuum
vector $\vacph\in\CH$ is stable under the action of the operators
$T(\al)$ and $T^{-1}(\al)$:
\beq\label{vac-inv}
T(\aa)\vacph=\sk{g'(\xi)}^{-1}\
\sum_{\ep=\pm}Z'_\ep(\aa)\cdot e^{\pi K}\cdot Z'_{-\ep}(\aa)\rav{Phidual}
e^{\pi K}\sum_{\ep=\pm}Z^{\prime*}_{-\ep}(\aa) Z'_{-\ep}(\aa)
\rav{complet}\vacph\ .
\eeq
Here and below we will often use the formulas
\beq\label{boost-act}
Z^*_\pm(\theta) e^{\pi K}= e^{\pi K}Z_\mp(\theta),\quad
Z'_\pm(\al) e^{\pi K}= e^{\pi K}Z^{\prime*}_\mp(\theta)
\eeq
which are consequences of the definition of the boost operator
and \r{Psidual}, \r{Phidual}.

The commutation relations \r{com-mon}
imply the commutativity
\beq\label{tr-mo-co}
[T(\aa_1),T(\aa_2)]=0
\eeq
which signifies that the operator $T(\al)$ can be considered as
the generating function of the local integrals of motion.
Using the property \r{ZFI-II} we can calculate the action of the generating
function $T(\al)$ onto $n$-particle state
 $|\theta_1,\ldots,\theta_n\rangle_{\ep_1,\ldots,\ep_n}$:
\beq\label{tr-act}
T(\al)\cdot |\theta_1,\ldots,\theta_n\rangle_{\ep_1,\ldots,\ep_n}
=\prod_{j=1}^n \ep_j\ \ctg\sk{\frac{\pi}{4}+\frac{\al-\theta_j}{2i}}
|\theta_1,\ldots,\theta_n\rangle_{\ep_1,\ldots,\ep_n}\ .
\eeq
Using this equality we can see that the quantity
\beq\label{gen-int}
I(\aa)=\sum_s I_se^{s\aa}=
\frac{1}{2i}\ T^{-1}(\aa)\frac{\partial T(\aa)}{\partial\aa}=
\frac{1}{2i}\ \frac{\partial\ln T(\al)}{\partial\al},
\eeq
has an eigenvalue on the states
$|\theta_N,\ldots,\theta_1\rangle_{\ep_N,\ldots,\ep_1}$
$$
I(\al)|\theta_N,\ldots,\theta_1\rangle_{\ep_N,\ldots,\ep_1}
=\sum_{j=1}^N\frac{2}{\ch(\al-\theta_j)}
|\theta_N,\ldots,\theta_1\rangle_{\ep_N,\ldots,\ep_1}
$$
and is a generating function of the local integrals of motion
\r{int-n} $I_n$ and $\bar I_n$ for odd indeces $n$:
\beq\label{int-act}
I(\al) |\theta_N,\ldots,\theta_1\rangle_{\ep_N,\ldots,\ep_1}=
\left\{
\begin{array}{c}
\sum_{s\geq0}(-1)^s e^{-(2s+1)\al}I_{2s+1}
|\theta_N,\ldots,\theta_1\rangle_{\ep_N,\ldots,\ep_1},\quad
\al\to+\infty \\
\sum_{s\geq0}(-1)^s e^{(2s+1)\al}\bar I_{2s+1}
|\theta_N,\ldots,\theta_1\rangle_{\ep_N,\ldots,\ep_1},\quad
\al\to-\infty
\end{array}
\right.
\eeq
where
\bea
I_{2s+1}
|\theta_N,\ldots,\theta_1\rangle_{\ep_N,\ldots,\ep_1}&=&
\sum_{j=1}^N e^{(2s+1)\theta_j}
|\theta_N,\ldots,\theta_1\rangle_{\ep_N,\ldots,\ep_1},\nn\\
\bar I_{2s+1}
|\theta_N,\ldots,\theta_1\rangle_{\ep_N,\ldots,\ep_1}&=&
\sum_{j=1}^N e^{-(2s+1)\theta_j}
|\theta_N,\ldots,\theta_1\rangle_{\ep_N,\ldots,\ep_1},\quad
s\geq0\ .\label{int-val}
\eea
It follows from \r{vac-inv} that
\beq\label{vacinv}
I_{2s+1}\vacph=\bar I_{2s+1}\vacph=0\ .
\eeq

It is clear that the form factors of the quantum integrals $I_{2s+1}$ and
$\bar I_{2s+1}$ vanish identically, but using these quantities we can
partially solve the problem of the reconstructing the map $O\to\tilde O$
of the local operators into the operators acting in the Hilbert space of the
angular quantization. Suppose we know this identification for some particular
local operator $\CO\to \tilde\CO$. Then we can immediately find
this identification for arbitrary descendant of the operator $\CO$
with respect to all integrals of motion  $I_{2s+1}$ and
$\bar I_{2s+1}$: $\CO(\al)=[\CO,I(\al)]$ \cite{Lu1}.
The answer is
\beq\label{ans11}
\CO(\al)\mapsto \tilde\CO(\al)=\tilde\CO\, \tilde I(\al)-
\tilde I(\al+2\pi i)\,\tilde \CO\ ,
\eeq
where
\beq\label{til-int}
\tilde I(\al)=\frac{1}{2ig'(\xi)}
\sum_{\ep=\pm}Z'_\ep(\al+i\pi)\partial_\al Z'_{-\ep}(\al)=
\sum_{s>0}\tilde I_{\pm(2s+1)}e^{\mp(2s+1)\al}\quad
\hbox{when}\quad \al\to\pm\infty\ .
\eeq
The prove of this statement is based on the cyclic property of the trace
\r{tracecon} and looks as follows:
\bea
&\lvacph \CO(\al) |\theta_1,\ldots,\theta_n\rangle_{\ep_1,\ldots,\ep_n}
=\ \ \lvacph
[\CO, I(\al)] |\theta_1,\ldots,\theta_n\rangle_{\ep_1,\ldots,\ep_n}
\nn\\
&\rav{vacinv}
\lvacph \CO I(\al) |\theta_1,\ldots,\theta_n\rangle_{\ep_1,\ldots,\ep_n}
=
\Tr_{\H}\sk{e^{2\pi K}\tilde\CO\tilde I(\al)
Z^*_{\ep_1}(\theta_1)\ldots Z^*_{\ep_n}(\theta_n)}-\nn\\
&-\frac{1}{2ig'(\xi)^2}
\sum_{\ep,\mu=\pm}(\ep\mu)^{n}
\Tr_{\H}\sk{e^{\pi K}\tilde \CO Z^{\prime*}_\ep(\al)Z'_\mu(\al)
Z^*_{\ep_1}(\theta_1)\ldots Z^*_{\ep_n}(\theta_n) e^{\pi K}
Z'_{-\mu}(\al)\partial_\al Z^{\prime*}_{-\ep}(\al)}.\nn
\eea
The
last line in the previous calculation  can be transformed as follows:
\bea
&\sum_{\ep,\mu=\pm}(\ep\mu)^{n}
\Tr_{\H}\sk{e^{\pi K}\tilde \CO Z^{\prime*}_\ep(\al)Z'_\mu(\al)
Z^*_{\ep_1}(\theta_1)\ldots Z^*_{\ep_n}(\theta_n) e^{\pi K}
Z'_{-\mu}(\al)\partial_\al Z^{\prime*}_{-\ep}(\al)}=\nn\\
&\quad\rav{ZFI-II}
\sum_{\ep,\mu=\pm}
\Tr_{\H}\sk{e^{\pi K} \tilde \CO
Z^*_{\ep_1}(\theta_1)\ldots Z^*_{\ep_n}(\theta_n)
 Z^{\prime*}_\ep(\al)
e^{\pi K} Z^{\prime*}_{-\mu}(\al)
Z'_{-\mu}(\al)\partial_\al Z^{\prime*}_{-\ep}(\al)}=\nn\\
&\quad\rav{complet}
g'(\xi)\sum_{\ep=\pm}
\Tr_{\H}\sk{ Z^{\prime*}_\ep(\al)
e^{\pi K} \partial_\al Z^{\prime*}_{-\ep}(\al)
e^{\pi K} \tilde \CO
Z^*_{\ep_1}(\theta_1)\ldots Z^*_{\ep_n}(\theta_n)
}=\nn\\
&\quad\rav{Phidual}
g'(\xi)\Tr_{\H}\sk{
e^{2\pi K} \tilde I(\al+2\pi i)  \tilde \CO
Z^*_{\ep_1}(\theta_1)\ldots Z^*_{\ep_n}(\theta_n)
}
\eea
so \r{ans11} is proved.

The generating function \r{til-int} has also another meaning.
Namely, it was proved in \cite{NPT} that
after substitution to the trace \r{tracecon}
the coefficients $\tilde I_{1}$ and
$\tilde I_{-1}$ of the asymptotical expansion \r{til-int}
one obtains the  known form factors
\cite{Sbook}
of the stress energy tensor in $SU(2)$-invariant Thirring model (this model
can be obtained from SG model in the limit $\xi\to+\infty$).
This allows to conjecture that for the finite $\xi$ the
corresponding coefficients of the quantity
$\tilde I(\al)$ will also generate the form factors of
stress-energy tensor in the SG model.

\subsection{Symmetries of the model}

In this subsection we will prove the following three statements.

($\iota$) The adjoint action of the algebra $\AA$ \r{adj}
on the total Hilbert space
 is given by the level zero
action of this algebra
\beq\label{repres1}
\overline \R(u_1-u_2,\xi)\Ad_{L_1(u_1;\xi)}\Ad_{L_2(u_2;\xi)}=
\Ad_{L_2(u_2;\xi)}\Ad_{L_1(u_1;\xi)} \overline \R(u_1-u_2,\xi)\ .
\eeq

($\iota\iota$) The subspace of the $n$-particle states carries the
finite-dimensional representation of the algebra $\AA$ given by the
the formulas
\beq\label{homoc}
\sk{\id\ot\iota\ot\id\ot\iota\ot\ldots\iota^{n-1}}\Delta^{(n-1)}(x)
|\theta_1,\ldots,
\theta_1\rangle_{\ep_1,\ldots,\ep_n},\quad
x=e(u),\ f(u),\ h(u)\ ,
\eeq
where
$\Delta^{(n)}(x)$ is $n$th power of the comultiplication maps
\r{com-e-fu}--\r{comul-h} defined inductively
$$\Delta^{(1)}\equiv\Delta,\quad
\Delta^{(n)}(x)=\sk{\Delta\ot\id}\Delta^{(n-1)}\ ,
$$
where the action of the Gauss coordinates
$e(u)$, $f(u)$ and $h(u)$ on the one-particle states is defined by
the formulas \r{e-res1} and \r{e-res2}.

($\iota\iota\iota$) The commutation relations of the algebra $\AA$ in the
form \r{ef}--\r{hh} allow to define certain asymptotical operators
$\ee_i$, $\ff_i$ and $\hh$ such that their commutation and
comultiplication relations correspond to those of the Chevalley generators
of the quantum affine algebra $U_q(\widehat{sl}_2)$ at level zero with
the parameter of deformation $q=\exp\sk{\pi i\frac{\xi+1}{\xi}}$.

The first statement is a simple consequence of the commutation relation
\r{scr-com} and the fact of commutativity $[\R(u,\xi),\sigma_z\otimes
\sigma_z]=0$.

The second statement is a direct consequence of the defining relations
\r{51a}, \r{52a} and \r{53c} for the operators $Z^*_\pm(\theta)$.
We start from one particle states $|\theta+\pi i/2\rangle_\pm$ and prove
that they realize the spin $1/2$ representation \r{eval-0}--\r{eval-h1}.
{}From the definition of the adjoint action \r{adj} we have
\bea
\Ad_{k_1(u)^{-1}}\cdot Z^*_\pm(\tilde\theta)&=&
k_1(\tilde u)^{-1}Z^*_\pm(\tilde\theta)k_1(\tilde u)+
k_1(\tilde u)^{-1}\{Z^*_\pm(\tilde\theta),f(\tilde u)\}k_2(\tilde u)
e(\tilde u)\label{ad11}\\
\Ad_{k_2(u)^{-1}}\cdot Z^*_\pm(\tilde\theta)&+&
\Ad_{e(u)k_1(u)^{-1}f(u)}\cdot Z^*_\pm(\tilde\theta)=\nn\\
&=&k_2(\tilde u)^{-1}Z^*_\pm(\tilde\theta)k_2(\tilde u)+
e(\tilde u)k_1(\tilde u)^{-1}\{Z^*_\pm(\tilde\theta),f(\tilde u)\}
k_2(\tilde u)\label{ad22}\\
-\Ad_{k_1(u)^{-1}f(u)}\cdot Z^*_\pm(\tilde\theta)&=&
k_1(\tilde u)^{-1}\{Z^*_\pm(\tilde\theta),f(\tilde u)\}
k_2(\tilde u)\label{ad12}\\
-\Ad_{e(u)k_1(u)^{-1}}\cdot Z^*_\pm(\tilde\theta)&=&
e(\tilde u)k_1(\tilde u)^{-1}Z^*_\pm(\tilde\theta)k_1(\tilde u)+
k_2(\tilde u)^{-1}Z^*_\pm(\tilde\theta)k_2(\tilde u)e(\tilde u)+\nn\\
&+&e(\tilde u)
k_1(\tilde u)^{-1}\{Z^*_\pm(\tilde\theta),f(\tilde u)\}
k_2(\tilde u)e(\tilde u)\label{ad21}
\eea
where we denote $\tilde u=u+\pi i/4$ and $\tilde \theta=\theta+\pi i/2$.

The calculation of the adjoint action of the Gauss coordinates of
$L$-operator onto the state $|\tilde \theta\rangle_-$ is an easy part.
Indeed, using formulas \r{53c} we observe first that the anticommutator
$\{Z^*_-(\tilde\theta),f(\tilde u)\}$ vanishes in \r{ad11}--\r{ad22}
and using then \r{51a}, \r{52a} we obtain
\beq\label{e-res1}
\Ad_{f(u)}|\tilde \theta\rangle_-=0, \quad
\Ad_{e(u)}|\tilde \theta\rangle_-=-\frac{\sh\,i\pi/\xi}
{\sh\, \sk{ \frac{u-\theta}{\xi} } }
|\tilde \theta\rangle_+,\quad
\Ad_{h(u)}|\tilde \theta\rangle_-=\frac{\sh\,\sk{\frac{u-\theta+i\pi}{\xi} }}
{\sh\,\sk{\frac{u-\theta}{\xi}}}|\tilde \theta\rangle_-
\eeq
which obviously coincide with the analogous formulas from
\r{eval-0}--\r{eval-h1}. Let us demonstrate how the second formula in
\r{e-res1} is obtained. Combining \r{ad11} and \r{ad21} and taking into
account \r{52c} we obtain
\bea
\Ad_{e(u)}\cdot Z_-^*(\tilde\theta)&=&
-e(\tilde u) Z_-^*(\tilde\theta) - h(\tilde u)Z_-^*(\tilde\theta)
h(\tilde u)^{-1}e(\tilde u)=\nn\\
&\rav{51a}&-e(\tilde u) Z_-^*(\tilde\theta) -
{\sh\frac{u-\theta+i\pi}{\xi}\over\sh\frac{u-\theta}{\xi}}Z_-^*(\tilde\theta)
e(\tilde u)=\nn\\
&\rav{52a}& -
{\sh\frac{i\pi}{\xi}\over\sh\frac{u-\theta}{\xi}}Z_+^*(\tilde\theta)\ .\nn
\eea

The calculation of the adjoint action of the Gauss
coordinates onto the state $|\tilde \theta\rangle_+$ is more
complicated but
straightforward. The main trick is to use the formula \r{52a} to replace
the operator $Z^*_+(\tilde\theta)$ by the combination of the products
$e(v)Z^*_-(\tilde\theta)$ and $Z^*_-(\tilde\theta)e(v)$. Using then the
commutation relations of the algebra $\AA$ in terms of the Gauss coordinates
\r{ef}--\r{hh} we will find that the dependence on the spectral parameter
$v$ is canceled out and we obtain
\beq\label{e-res2}
\Ad_{e(u)}|\tilde \theta\rangle_+=0, \quad
\Ad_{f(u)}|\tilde \theta\rangle_+=-\frac{\sh\,i\pi/\xi}
{\sh\,\sk{\frac{u-\theta}{\xi}}}|\tilde \theta\rangle_-,\quad
\Ad_{h(u)}|\tilde
\theta\rangle_+=\frac{\sh\,\sk{\frac{u-\theta-i\pi}{\xi}}}
{\sh\,\sk{\frac{u-\theta}{\xi}}}|\tilde \theta\rangle_+
\eeq
which coincide with the rest of the formulas \r{eval-0}--\r{eval-h1}.

To find the action of the Gauss coordinates $e(u)$, $f(u)$ and $h(u)$
on the $n$-particle states we use the same formulas \r{ad11}--\r{ad21}
with $Z^*_\pm(\tilde\theta)$  replaced by the
$n$-fold product of these operators.
For example, the adjoint action of the Gauss coordinate $e(u)$ on the
two-particle state is given by the formula
\bea
  \Ad_{e(u)}\cdot|\theta_1,\theta_2\rangle_{-,-}&=&
-{\sh\sk{\frac{i\pi}{\xi}}\over\sh\sk{\frac{u-\theta_1}{\xi}}}
|\theta_1,\theta_2\rangle_{+,-}
+
{\sh\sk{\frac{u-\theta_1+i\pi}{\xi}}\over\sh\sk{\frac{u-\theta_1}{\xi}}}
{\sh\sk{\frac{i\pi}{\xi}}\over\sh\sk{\frac{u-\theta_2}{\xi}}}
|\theta_1,\theta_2\rangle_{-,+}=\nn\\
&=&\hat\Delta(e(u))\ |\theta_1,\theta_2\rangle_{-,-} \ ,
\label{examp1}
\eea
where we denoted by $\hat\Delta=\sk{\id\ot\iota}\Delta$
the composition of the comultiplication of the algebra $\AA$
\r{com-e-fu}--\r{comul-h} and the involution \r{L-oper-rel}.
Repeating these arguments inductively
we prove the formula \r{homoc} where
the action of the Gauss coordinates $e(u)$, $f(u)$ and $h(u)$
on the one-particle states are given by the formulas
\r{e-res1} and \r{e-res2}.

The commutation relations of the algebra $\AA$
\r{ef}--\r{hh}
at the zero central element
demonstrate that the Gauss coordinates of $L$-operators
have following asymptotics when $\Re\,u\to\pm\infty$:
\beq
e(u)\stackreb{\sim}{}
\exp\sk{-\frac{|u|}{\xi}},\quad
f(u)\stackreb{\sim}{}
\exp\sk{-\frac{|u|}{\xi}},\quad
h(u)\stackreb{\sim}{}  h(\pm\infty)\equiv h_\pm\ .
\label{asympHH}
\eeq

It follows from \r{he} and \r{hf} that
Cartan asymptotical generators $h_\pm$ have the following commutation
relations with Gauss coordinates $e(u)$ and $f(u)$:
\beq
h_\pm e(u) h_\pm^{-1}=\exp\sk{\pm\frac{2\pi i}{\xi}}e(u),\quad
h_\pm f(u) h_\pm^{-1}=\exp\sk{\mp\frac{2\pi i}{\xi}}f(u)
\ .
\label{q-comm}\eeq
The comultiplication rule \r{comul-h} yields that the asympotical
Cartan elements
are primitive and  group-like: $\Delta h_\pm=h_\pm\otimes h_\pm$.
The commutation relations \r{q-comm} yields that
the product  $h_+ h_-$
is central and also  group-like primitive.
Due to this we can put this central element
 to be equal to one so the asymptotical
Cartan operators are inverse to each other: $h_+=h_-^{-1}$.

Let us define
the logarithmic Cartan operator $\hh$ as follows:
\beq\label{as-Car}
h_\pm=\exp\sk{\pm i\pi \frac{\xi+1}{\xi}\ \hh}
\eeq
where the operator $\hh$ has standard commutation relation with the Gauss
coordinates $[\hh,e(u)]=2e(u)$ and $[\hh,f(u)]=-2f(u)$.
Define also the asymptotical operators
\bea
  \ee_\pm&=&\frac{1}{2}\
\sh\sk{i\pi\frac{\xi+1}{\xi}}^{-1}
\lim{\Re\,u\to\pm\infty}e^{\pm u/\xi}e(u)\ ,\nn\\
 \ff_\pm&=&\frac{1}{2}\ \sh\sk{i\pi\frac{\xi+1}{\xi}}^{-1}
\lim{\Re\,u\to\pm\infty}e^{\pm u/\xi}f(u)\ .
\label{asy-gen}
\eea
{}From the commutation relations \r{ef}--\r{hf} we can obtain the
commutation relations of these operators:
\bea
[\hh,\ee_\pm]&=&\pm2\ee_\pm,\qquad [\hh,\ff_\pm]=\mp2\ff_\pm,\qquad
[\ee_\pm,\ff_\mp]=\pm
\frac{\sin\sk{\pi \hh(\xi+1)/\xi}}{\sin\sk{\pi(\xi+1)/\xi}}=
\pm\frac{q^\hh-q^{-\hh}}{q-q^{-1}}\ ,\nn\\
q\ee_+\ee_-&=&{q^{-1}}\ee_-\ee_+,\quad
q^{-1}\ff_+\ff_-\ =\ {q}\ff_-\ff_+\ ,\nn\\
q^{\mp3}\ee^3_\pm\ff_\pm&-&\frac{q^3-q^{-3}}{q-q^{-1}}
\sk{q^{\mp1}\ee^2_\pm\ff_\pm\ee_\pm
-q^{\pm1}\ee_\pm\ff_\pm\ee^2_\pm}-q^{\pm3}\ff_\pm\ee^3_\pm=0\ ,\nn\\
q^{\pm3}\ff^3_\pm\ee_\pm&-&\frac{q^3-q^{-3}}{q-q^{-1}}
\sk{q^{\pm1}\ff^2_\pm\ee_\pm\ff_\pm
-q^{\mp1}\ff_\pm\ee_\pm\ff^2_\pm}-q^{\mp3}\ee_\pm\ff^3_\pm=0\ ,
\label{soot1}
\eea
where $q=\exp\sk{i\pi\frac{\xi+1}{\xi}}$.
These commutation relations allows to identify the asymptotical operators
with the Chevalley generators of the
affine quantum algebra  $U_q(\widehat{sl}_2)$ at level zero.

Using formulas \r{e-res1}, \r{e-res2} and the rule of the Gauss coordinates
actions onto multi-particle states \r{homoc}
we can obtain
the action of the asymptotical operators
$\ee_\pm$, $\ff_\pm$ and $\hh$ onto
multiparticle states and prove that it is given
by the comultiplication of the algebra $U_q(\widehat{sl}_2)$.
 To do this we first slightly modify
the action of these generators following \cite{RS} when they act on the
one-particle states $|\theta\rangle_\pm$:
\begin{equation}\label{as-ge-mo}
  \ee_\pm\mapsto \exp\sk{\mp\frac{\theta}{\xi}} \ee_\pm,\quad
 \ff_\pm\mapsto \exp\sk{\pm\frac{\theta}{\xi}} \ff_\pm,\quad \hh \mapsto \hh.
\end{equation}

By the straightforward calculation using the definition of the adjoint
action on the multiple-particle states \r{adj}
and the formulas \r{51a}--\r{53c} we obtain that this action can be
formulated through the comultiplication
\begin{equation}\label{co-as}
  \Delta_0 \ee_\pm= \ee_\pm\ot 1+q^{\mp\hh}\ot \ee_\pm,\quad
 \Delta_0 \ff_\pm= 1\ot \ff_\pm + \ff_\pm\ot q^{\pm\hh},\quad
\Delta_0 \hh=\hh\ot 1+1\ot \hh\ ,
\end{equation}
which can be formally obtained from the comultiplication formulas for
the Gauss coordinates \r{com-e-fu}--\r{comul-h} using \r{asympHH}.
The action of the asymptotical operators on the one particle states are
defined as follows
\beq\label{act-asy}
\ee_\pm|\theta\rangle_+=\ff_\pm|\theta\rangle_-=0,\quad
\ee_\pm|\theta\rangle_-=|\theta\rangle_+,\quad
\ff_\pm|\theta\rangle_+=|\theta\rangle_-,\quad
\hh|\theta\rangle_\pm=\pm|\theta\rangle_\pm\ .
\eeq
For example, let us demonstrate the origin of this comultiplication on
the two-particle state. From \r{examp1} we have
\bea
\Ad_{\ee_+}\cdot|\theta_1,\theta_2\rangle_{-,-}\!\!\!\!&=&\!\!\!\!\!\!\!\!
\lim{\Re u\to+\infty}\sk{{\exp\sk{\frac{u-\theta_1}{\xi}}\over
2\sh\sk{\frac{u-\theta_1}{\xi}}}|\theta_1,\theta_2\rangle_{+,-}
-
{\sh\sk{\frac{u-\theta_1+i\pi}{\xi}}\over\sh\sk{\frac{u-\theta_1}{\xi}}}
{\exp\sk{\frac{u-\theta_2}{\xi}}\over2\sh\sk{\frac{u-\theta_2}{\xi}}}
|\theta_1,\theta_2\rangle_{-,+}}=\nn\\
&=&|\theta_1,\theta_2\rangle_{+,-}-\exp\sk{\frac{i\pi}{\xi}}
|\theta_1,\theta_2\rangle_{-,+}=\nn\\
&\rav{act-asy}&\sk{\ee_+\ot1+q^{-\hh}\ot\ee_+}|\theta_1,\theta_2\rangle_{-,-}
=\Delta_0(\ee_+)\ |\theta_1,\theta_2\rangle_{-,-}\ .\nn
\eea
We would like to note here that the set of the asymptotical generators
$\ee_+$, $\ff_-$ and $\hh$
or $\ee_-$, $\ff_+$ and $\hh$
 cannot be identified with the set $e$, $f$ and $h$
used in the construction of the evaluation homomorphism from the algebra
$\AA$ onto $U_{i\pi/\xi}(sl_2)$, because the first ones are the
 subalgebras while
the second one is factor subalgebra. In particular,
the action \r{act-asy} cannot be obtained from the adjoint action
onto one-particle states \r{e-res1} and \r{e-res2}.

The consideration presented above  prove
 that the adjoint action of the finite-dimensional
subalgebra of $\AA$ onto the
total Hilbert space of the SG model describe the symmetries of this space
investigated in \cite{RS,AL,BL}.

\subsection{Symmetries of the model at the FF point}

Now we would like
to demonstrate how the quantum symmetries of the Hilbert space
$\CH$ of the SG model become the
classical ones (i.e.
correspond to undeformed current algebra) at the FF point.

It is clear that the finite-dimensional representations of the algebra
$\AA$ at the value  $\xi=1$ degenerate. Moreover,
the operator $h_+$ becomes the central element of the algebra (cf.
\r{q-comm}) and takes the value
$(-1)^{k+1}$, $k=0,1$ on the subspace $\CH_k$ of the even and odd
number of particles of the total
Hilbert space $\CH$.
In order to obtain the nontrivial action
of the algebra $\AA$ at the FF point on the Hilbert space
of states we introduce the rescaled operators
\beq\label{res-op}
\hat e(u)=\left.-\frac{e(u)}{\sh(i\pi/\xi)}\right|_{\xi=1},\quad
\hat f(u)=\left.-h_+\frac{f(u)}{\sh(i\pi/\xi)}\right|_{\xi=1},\quad
\hat h(u)=\left.-\frac{h(u)h_+^{-1}-1}{\sh(i\pi/\xi)}\right|_{\xi=1}.
\eeq
The nontrivial
commutation relations of the algebra $\AA$ reads as follows:
\bea
{[}\hat h(u), \hat e(v){]}&=& 2\cth\,(u-v) \hat e(v)-
2{\hat e(u)\over \sh\,(u-v)} \ ,\label{the}\\
{[}\hat h(u), \hat f(v){]}&=& -2\cth\,(u-v) \hat f(v)+
2{\hat f(u)\over \sh\,(u-v)} \ ,\label{thf}\\
{[}\hat e(u), \hat f(v){]}&=& {\hat h(u)-\hat h(v)
\over  \sh\,(u-v) }\  .\label{thh}
\eea
The algebra \r{the}--\r{thh} coincides with the classical current algebra
$\widehat{sl}_2$
on the line \cite{KLPST}.

Formulas \r{e-res1} and \r{e-res2} of the adjoint action of
the operators \r{res-op} becomes
\bea
\Ad_{\hat f(u)}|\tilde \theta\rangle_-&=&
\Ad_{\hat e(u)}|\tilde \theta\rangle_+=0, \quad
\Ad_{\hat h(u)}|\tilde \theta\rangle_\pm=\pm\cth(u-\theta)
\theta\rangle_\pm\ ,\nn\\
\Ad_{\hat e(u)}|\tilde \theta\rangle_-&=&\frac{1} {\sh\, \sk{u-\theta} }
|\tilde \theta\rangle_+,\quad
\Ad_{\hat f(u)}|\tilde \theta\rangle_+=-\frac{1}
{\sh\,\sk{u-\theta}}|\tilde \theta\rangle_-\ ,\label{e-res0}
\eea
and on the multi-particle states are
\beq
\hat\Delta(x)=x\ot1+1\ot x,\quad x=\hat e(u),\ \hat f(u),\ \hat h(u)\ ,
\label{com-clas}
\eeq
where in order to obtain \r{e-res0} and \r{com-clas}
we used the fact that  operator $h_+$ equal to $-1$ on the
one-particle state.

The phenomena that quantum symmetries of the Hilbert space of state for
the SG model becomes the classical ones at the FF point
is a consequence of the fact that $S$-matrix in this limit yields
the classical $r$-matrix \r{r-clas}:
\beq\label{r-clas2}
r(u)=\lim{\xi\to1}\frac{S(u;\xi)+1}{\pi i(1-\xi)}\ .
\eeq
This phenomena was observed in
reflectionless
SG theory
\cite{LN} and was used
to investigate the space of the local operators in
SG model at FF point \cite{Le}.

\setcounter{equation}{0}
\section{Discussion}

In this paper we further developed  the method of angular quantization
for the Sine-Gordon model. Technically the application of this method splits
into two parts. First, one should explicitly describe canonical quantization
$\CH_R$ of the model in right Rindler wedge, where the boost plays the role
of hamiltonian. Then the space of states and local operators of the theory
on the line are described in terms of certain operators acting in $\CH_R$.

We studied the SG theory at the free fermion point where the canonical
quantization in RRW can be done explicitly. We investigated the integrals of
motion and found that
 the usual local integrals of motion diverge. This forced us to
consider nonlocal integrals of motion which are a certain
analytical continuation  (in the space of eigenvalues of Lorentz boost) of
the usual charges and the only possibility to close them into a quadratic
current algebra is to use charges with different monodromy properties.
They form the specialization of the scaling elliptic algebra
$\AA$  \cite{KLP}
into free fermion point. The bosonization \cite{Lu1} naturally appears in
terms of scattering data.

This indicates that angular quantization of  SG model can be done in terms of
the representation theory of the algebra $\AA$.
Starting from level one representations of this algebra in the bosonic Fock
space we managed to construct the space of
asymptotical states of SG model and some local
operators acting into this space of states, in particular, the
transfer matrix and the
commuting set of the integrals of motion,  and demonstrate the mechanism of
trace calculations of the form factors of local operators.
This approach is an extension  of the ideas presented in \cite{JM} for
XXZ model. The algebra $\AA$ is not a Hopf algebra, but we were able to
define the adjoint action of this algebra on the space of states, such
that $n$-particle states with given rapidities form $n$-fold tensor product
of two-dimensional representations of the algebra $\AA$.

Contrary to the integrable models on the lattice local integrals and local
operators of the SG theory appear as coefficients of the asymptotical
expansions of certain currents which are constructed explicitly. In
particular, the asymptotical expansion of the level zero adjoint action of
the algebra $\AA$ on the space of states produce the action of Chevelley
generators of quantum affine algebra, which was known before. At the free
fermion point we get in this way the action of the classical affine algebra
$\widehat{sl}_2$ which was constructed in the framework of the radial
quantization in the paper \cite{Le}.

Nevertheless, the understanding of the angular quantization method of SG
model for generic value of the renormalized coupling constant $\xi$ is
far from being complete. In particular, there is no rigorous construction of
the quantum analogs of the Jost functions introduced in \cite{Lu1,Lu2}
without referring to bosonization.
SG model admits also natural analog of `new level zero action'
(see \cite{JMKKP} and references therein) which is given in terms of
$L$-operators as follows
\beq\label{adj-d}
\Ad'_{L(u)}\cdot X= L(u)\,X\,L(u)^{-1}\
\eeq
and depends on the dual deformation parameter
$q'=\exp\sk{i\pi\frac{\xi}{\xi+1}}$.
It will be interesting to extend the results on the spinon bases
 in conformal field theories investigated in \cite{Spinon}
to the massive integrable models.
As we mentioned already that the algebra $\AA$ is  quasi-Hopf
algebra, but belonging to a family of dynamical elliptic algebra.
 The definition of adjoint action, used in this paper, did not
 refer to the axiomatics of this family. It would be interesting to
 fill this gap.

Finally, it would also be interesting to further explore the role
of the duality transformation
\r{dual-tr}.  In terms of the SG coupling $\beta$, this is an
electric/magnetic duality $\beta \to 2/\beta$ familiar from the
conformal field theory of a compactified free boson.  As described in the
paper this duality relates the $q'$-deformation parameter of
the algebra of the monodromy matrix with the $q$-deformation parameter
of the physical S-matrix.  Though we did not present this here,
one can  define a dual monodromy matrix by the
formal replacement $\beta \to 2/\beta$ in the usual monondromy matrix,
and show that formally this dual monodromy matrix commutes with the
original monodromy matrix.  This would imply that the dual monodromy
matrix generates additional integrals of motion, presumably related
to the quantum affine symmetry described in \cite{BL}.

\section{Acknowledgment}

This investigation was basically done  during the visits of
two of the authors
(S.Kh. and S.P.) to Cornell University organized in the framework
of the CRDF Award  No. RM2-150.
We would like to acknowledge  hospitality at the Newmann Laboratory
of Cornell University and to thank S.~Lukyanov and A.~Zamolodchikov for the
hospitality and useful discussions during S.Kh. and S.P. visit to
Rutgers University in November 1998. We also would like to acknowledge
the discussions with our colleagues S.~Kharchev, D.~Lebedev and V.~Shadura.

The research described in this publication was made possible in part by
grants
RFBR-97-01-01041 (S.~Pakuliak),
INTAS 93-10183 (S.~Khoroshkin),
INTAS--OPEN--97--1312,
by Award  No. RM2-150 of the U.S. Civilian Research \& Development
Foundation (CRDF) for the Independent States of the Former Soviet Union,
and
by the grants for promotion of French-Russian
scientific cooperation: the CNRS grant PICS No. 608 and the RFBR grant No.
98-01-2033 (S.~Khoroshkin and S.~Pakuliak).

\app{The algebra of bosons $a_\la$ and $\tilde a_\la$}
\label{app1}

The normal ordering with respect to the vacuum vectors
 \r{Fock} and \r{lFock}
\beq
\label{j7}
b_\pm(\nu)b_\mp(\nu')={:}b_\pm(\nu)b_\mp(\nu'){:} +
\langle b_\pm(\nu)b_\mp(\nu')\rangle,\qquad
\langle b_\pm(\nu)b_\mp(\nu')\rangle = \delta(\nu+\nu')\Theta(-\nu'),
\eeq
where $\Theta(\nu)$ is a `continuous'  step function
\beq\label{step}
\Theta(\nu)=\left\{
\begin{array}{c}
1,\quad\hbox{for}\quad \nu>0\\
1/2,\quad\hbox{for}\quad \nu=0\\
0,\quad\hbox{for}\quad \nu<0
\end{array}\right.,\qquad
\Theta(\nu)+\Theta(-\nu)\equiv1
\eeq
allows to observe that  the commutation relations
between operators $a_\la$ and $\tilde a_\la$ is not closed in a sense  that
the commutator $[a_\la,\tilde a_\mu]$ cannot be presented as a linear
combinations of the same operators with $\CC$-number coefficients
and $\CC$-valued functions. The idea is to consider this commutator
as a new bosonic operator and try to close the extended by this operator
algebra. Fortunately this extended algebra is closed. To describe its
commutation relations we introduce new operators:
\beq\label{j14a}
t_{\la,\mu}=
\intt d\nu\ \frac{\Gamma\sk{\frac{1}{2}-i(\la+\nu)}}
{\Gamma\sk{\frac{1}{2}+i(\mu-\nu)}}\ \th\,\pi\nu\
{:}b_-(\mu-\nu)b_+(\la+\nu){:}\ ,
\eeq
\beq\label{j14}
\tilde t_{\la,\mu}=
\intt d\nu\ \frac{\Gamma\sk{\frac{1}{2}-i(\mu-\nu)}}
{\Gamma\sk{\frac{1}{2}+i(\la+\nu)}}\ \th\,\pi\nu\
{:}b_-(\mu-\nu)b_+(\la+\nu){:}\ .
\eeq
The bosonic operators
$a_\la$ and $\tilde a_\la$ are related to
the new operators $t_{\la,\mu}$ and $\tilde t_{\la,\mu}$
by the linear transformation:
\bea
\tilde a_\la&=&\cth\,\pi\la\ \tilde t_{\la,0} -
\frac{1}{\sh\,\pi\la}\  t_{0,\la},\nn\\
a_\la&=& \frac{1}{\sh\,\pi\la}\ \tilde t_{\la,0} -\cth\,\pi\la\
t _{0,\la}. \label{j17}
\eea

Using simple trigonometric algebra we conclude that the set of the
operators $t_{\la,\mu}$ and $\tilde t_{\la,\mu}$ is not independent.
For example, the following relation is valid:
\beq\label{j19}
\sh\,\pi\la\ (t_{\la,\mu}- t_{\la+\mu,0}
=
\sh\,\pi\mu\ (\tilde t_{0,\la+\mu}-\tilde t_{\la,\mu})
\eeq
therefore we can conclude that complete
algebra of the bosonic operators  reads as follows:
\bea
[a_\la,a_\mu{]}&=&\la\delta(\la+\mu)\label{j21}\\
{[}a_\la,t_{\mu,\rho}{]}&=&
t_{\mu,\rho+\la}-t_{\mu+\la,\rho}+\delta(\la+\mu+\rho)
\int_{0}^\lambda d\nu\ \th\,\pi(\nu+\rho)\label{j15}\\
{[}t_{\la,\mu},t_{\la',\mu'}{]}
&=&
\cth\,\pi(\la+\mu')(t_{\la+\la'+\mu',\mu}-t_{\la',\la+\mu'+\mu})+\nn\\
&+&
\cth\,\pi(\la'+\mu)(t_{\la,\la'+\mu'+\mu}-t_{\la'+\la+\mu,\mu'})+
\nn\\
&+&\delta(\la+\la'+\mu+\mu')\int_0^{\la+\mu}
d\nu \ \th\,\pi(\nu-\la)\ \th\,\pi(\nu+\mu').
\label{j16}
\eea
This algebra can be understood as an algebraic realization of the
complicated integral transform which relate the operators $\tilde a_\la$ and
$a_\la$. Indeed, using these commutation relations we can verify that
the combination
\beq\label{j20}
\tilde a_\la = \ch\,\pi\la\ a_\la - \sh\,\pi\la\ t_{0,\la}
\eeq
also has the commutation relations of the Heisenberg algebra
as $a_\la$ do (cf. \r{j21}).

\app{Quantum Jost functions at the FF point}
\label{app2}

We will prove the equivalence of
 \r{rel+}  to
\r{rel1}. The second case can be treated analogously.

First, we write down explicitly all the normal ordering rules
which follows from \r{normal}
\bea
Z'_-(z)F(u)&=&\frac{e^{\gamma/2}}{(2\pi)^{3/2}}
{
\Ga{\frac{1}{4}+\frac{u-z}{2\pi i}}
 \over
\Ga{\frac{3}{4}+\frac{u-z}{2\pi i}}
 }\ {:}Z'_-(z)F(u){:}\nn\\
F(u)Z'_-(z)&=&\frac{e^{\gamma/2}}{(2\pi)^{3/2}}
{
\Ga{\frac{1}{4}-\frac{u-z}{2\pi i}}
 \over
\Ga{\frac{3}{4}-\frac{u-z}{2\pi i}}
 }\ {:}Z'_-(z)F(u){:}\nn\\
F(u)F(v)&=& -\frac{e^{2\gamma}}{2\pi i}(u-v)
{:}F(u)F(v){:}\nn\\
Z'_+\sk{\al_1-\frac{i\pi}{2}}
Z'_+\sk{\al_2+\frac{i\pi}{2}}&=&g(\al_1-\al_2){:}Z'_+\sk{\al_1-\frac{i\pi}{2}}
Z'_+\sk{\al_2+\frac{i\pi}{2}}{:}\nn
\eea
where in the last formula the function $g(\al)$ is given in terms of double
$\Gamma$-functions
$$
g(\al)=\frac{e^{5\gamma/4}}{2\pi}
\frac{\Gamma_2^2(3\pi+i\al)}{\Gamma_2(2\pi+i\al)\Gamma_2(4\pi+i\al)}.
$$
For the double and usual $\Gamma$-functions we use the integral
representations
\cite{Bar,JM1}
\bea
\int_{\tilde C}{d\la\,\ln(-\la)\over2\pi i\la}\
{e^{-x\la}\over1-e^{-\la/\eta}}&=&\ln\Gamma(\eta x)+
\left(\eta x-\fract{1}{2}\right)(\gamma-\ln\eta)-\fract{1}{2}\ln2\pi\ ,
\nn
\\
\int_{\tilde C}{d\la\,\ln(-\la)\over2\pi i\la}\
{e^{-x\la}\over(1-e^{-\la\omega_1})
(1-e^{-\la\omega_2})}&=&\ln\Gamma_2(x\mid\omega_1,\omega_2)-{\gamma\over2}
B_{2,2}(x\mid\omega_1;\omega_2)\ ,
\nn
\eea
where $B_{2,2}(x\mid\omega_1;\omega_2)$ is the double Bernulli polynomial
of the second order
$$
B_{2,2}(x\mid\omega_1;\omega_2) ={1\over\omega_1\omega_2}
\left[x^2-x(\omega_1+\omega_2)+{\omega^2_1+3\omega_1\omega_2+\omega^2_2
\over6}\right].
$$
The constant $g$ in the relation \r{rel1} is the value of the function
$g(\al)$ at the point $\al=0$.

We have
\bea
&g^{-1}
Z'_+\sk{\al-\frac{i\pi}{2}}
Z'_+\sk{\al+\frac{i\pi}{2}}
=\nn\\
&=\frac{i}{16\pi^6}\int_{C_1}\int_{C_2} du_1du_2\
{:}
Z'_-\sk{\al-\frac{i\pi}{2}}
Z'_-\sk{\al+\frac{i\pi}{2}}
F(u_1)F(u_2){:}\times\nn\\
&\quad\times
\Ga{\frac{1}{2}+\frac{u_1-\al}{2\pi i}}
\Ga{-\frac{u_1-\al}{2\pi i}}
\Ga{\frac{1}{2}-\frac{u_2-\al}{2\pi i}}
\Ga{\frac{u_2-\al}{2\pi i}} \times\nn\\
&\quad  \times (u_1-u_2)\
{
\Ga{\frac{1}{2}+\frac{u_2-\al}{2\pi i}}
\Ga{\frac{1}{2}-\frac{u_1-\al}{2\pi i}}
 \over
\Ga{1+\frac{u_2-\al}{2\pi i}}
\Ga{1-\frac{u_1-\al}{2\pi i}}}.\label{r4}
\eea
The contours $C_1$ and $C_2$ in \r{r4} go from $-\infty$ to $+\infty$ and
\beq\label{co1}
\Im\,\al-\pi      < \Im\,u_1<  \Im\,\al ,\quad
\Im\,\al      < \Im\,u_2<  \Im\,\al+\pi
\eeq
Using the elementary properties of the $\Gamma$-functions
we can rewrite the integrand in \r{r4} in the form
\beq\frac{i}{8\pi^2}
\int_{C_1}\int_{C_2} du_1du_2
\left[{1\over u_2-\al}
-{1\over u_1-\al}\right]
{{:}
Z'_-\sk{\al-\frac{i\pi}{2}}
Z'_-\sk{\al+\frac{i\pi}{2}}
F(u_1)F(u_2){:}\over
\ch\sk{\frac{u_1-z}{2}}
\ch\sk{\frac{u_2-z}{2}}
}\label{integra}\eeq
where contours $C_1$ and $C_2$ are specified in \r{co1}.

Using the fact that  integrand in \r{integra} is antisymmetric function with
respect to variables $u_1$ and $u_2$ we conclude:
\beq\label{final}
g^{-1}
Z'_+\sk{\al-\frac{i\pi}{2}}
Z'_+\sk{\al+\frac{i\pi}{2}}=\frac{1}{4\pi}\intt du \frac{F(u)}
{\ch\sk{\frac{u-\al}{2}}}=\Lambda_-(\al)
\eeq
since the current $F(u)$ coincide with the scattering data operator
$\Z_-(u)$.

\end{document}